\documentclass[a4paper,12pt]{article}
\pdfoutput=1
\usepackage{newtxmath,newtxtext}
\usepackage{graphicx}
\usepackage{caption,subcaption}
\usepackage[usenames,dvipsnames]{xcolor}
\usepackage{bm}
\pdfoutput = 1
\usepackage{makecell}
\usepackage{hyperref}
\hypersetup{
	colorlinks=true,
	linkcolor=blue,     
	urlcolor=black,
	citecolor=blue
}

\usepackage{epsfig}%
\title{\centerline \bf Dynamical Stability in presence of non-minimal derivative dependent coupling of $ k $-essence field with a relativistic fluid}
\bigskip
\author{Kaushik Bhattacharya$^\dagger$, Anirban Chatterjee $^\star$,  Saddam Hussain $^\$ $  
	\thanks{ $^\dagger$kaushikb@iitk.ac.in,$^\star$anirbanc@iitk.ac.in, $^\$$msaddam@iitk.ac.in }
	\\
	\normalsize
	Department of Physics, Indian Institute of Technology, Kanpur\\ 
	\normalsize
	Kanpur 208016, India
}

\newcommand{\ti}[1]{\ensuremath{ \tilde{#1}}}
\newcommand{\nb}{\ensuremath{ \nabla }}

\newcommand{\m}{\ensuremath{{\mu \nu}}}
\newcommand{\la}{\ensuremath{{\lambda \nu}}}
\newcommand{\p}{\ensuremath{\partial{}}}
\newcommand{\eqq}[1]{Eq.~\eqref{#1}}
\begin{document}

\maketitle

\begin{abstract}
  In this paper we investigate a non-minimal, space-time derivative
  dependent, coupling between the $k$-essence field and a
  relativistic fluid using a variational approach. The derivative
  coupling term couples the space-time derivative of the $k$-essence
  field with the fluid 4-velocity via an inner product. The inner
  product has a coefficient whose form specifies the various models of
  interaction.  By introducing a coupling term at the Lagrangian level
  and using the variational technique we obtain the $k$-essence field
  equation and the Friedmann equations in the background of a
  spatially flat Friedmann-Lemaitre-Robertson-Walker (FLRW)
  metric. Explicitly using the dynamical analysis approach we analyze
  the dynamics of this coupled scenario in the context of two kinds of
  interaction models.  The models are distinguished by the form of the
  coefficient multiplying the derivative coupling term. In the simplest
  approach we work with an inverse square law potential of the
  $k$-essence field. Both of the models are not only capable of
  producing a stable accelerating solution, they can also explain
  different phases of the evolutionary universe.
\end{abstract}
\section{Introduction}

We have entered the age of precision cosmology. Late time cosmic acceleration is one of the important discoveries of this era. Various observational artifacts like measurement of luminosity distance and red-shift of type Ia supernova \cite{Riess:1998cb}, baryon acoustic oscillation \cite{Eisenstein:2005su}, Cosmic microwave background radiation \cite{Gawiser:2000az} power spectrum show that the present universe is going through a phase of accelerated expansion. It is widely assumed that an esoteric entity called `Dark energy,' is solely responsible for this late time acceleration. The origin and constituents of this entity are still a big mystery. To tackle the question related to the cause of late time acceleration the $\Lambda$-CDM \cite{Carroll:1991mt} model has been introduced, where $\Lambda$ corresponds to the cosmological constant and CDM refers to the cold dark matter. Despite the countless success, $\Lambda$-CDM suffers from fine-tuning and cosmic coincidence problems. Another important downside of the $\Lambda$-CDM model is, from the observational perspective, the discrepancy between the present observed value of Hubble's constant and the predicted value of Hubble's constant from theory. This statistically significant tension has arisen due to the estimation of the Hubble constant, $H_0$, from the Cosmic Microwave Background (CMB) and the direct local distance ladder measurements. In $\Lambda$-CDM scenario, analyzed data of Planck-CMB gives $H_0 = 67.4 \pm 0.5  \rm km s^{-1} Mpc^{-1} $, which is in  $4.4\sigma$ tension with the local measurement $H_0 =74.03 \pm 1.42 \rm km s^{-1} Mpc^{-1}$ \cite{DiValentino:2021izs, Lucca:2020zjb, Salvatelli:2013wra}. In addition to that, several other observations also suggest a higher value of the Hubble constant. These fundamental discrepancies motivate us to study various non-minimally coupled field-fluid models of dark sector.

Several theoretical ventures of field-theoretic models have already been introduced to investigate the behavior of dark energy. Among them, canonical \cite{Peccei:1987mm,Ford:1987de,Peebles:2002gy,Nishioka:1992sg, Ferreira:1997au,Ferreira:1997hj,Caldwell:1997ii,Carroll:1998zi,Copeland:1997et,Zlatev:1998tr,Hebecker:2000au,Hebecker:2000zb}, non-canonical \cite{Fang:2014qga,ArmendarizPicon:1999rj,ArmendarizPicon:2000ah,ArmendarizPicon:2000dh,ArmendarizPicon:2005nz,Chiba:1999ka,Chiba:2002mw,Kang:2007vs,ArkaniHamed:2003uy, Caldwell:1999ew} scalar field formulations, $f(R)$ modified gravity theory \cite{Capozziello:2002rd, Carroll:2003wy} and scalar-tensor theory \cite{Amendola:1999qq, Uzan:1999ch} are well known. Scenarios with a canonical scalar field and a relativistic fluid, with algebraic and derivative coupling terms, have already been explored in several pieces of literature \cite{Brown:1992kc,Boehmer:2015kta,Boehmer:2015sha,Shahalam:2017fqt,Barros:2019rdv,Kerachian:2019tar, Gumjudpai:2015vio,Cai:2017dxl,Mostaghel:2016lcd}. Advancing this scenario, we are introducing a non-canonical scalar field with a relativistic fluid which are interacting with each other non-minimally. In the present paper the non-minimal field-fluid coupling involves the single order derivative of the $k-$essence scalar field. In the present paper the field sector is governed by a non-canonical $k$-essence scalar field. The fluid sector is comprised by an ideal relativistic hydrodynamic fluid. Using a variational approach, we will analyze this coupled system in a flat homogeneous Friedmann-Lemaitre-Robertson-Walker universe. 

In this paper, we have taken the arbitrary form of the interaction between field-fluid as $f(n,s,\phi,X) U^{\mu} \p_{\mu} \phi$. The interaction term $f$ contains a non-canonical kinetic term $ X $, which contains the spacetime derivative of the scalar field. Derivative of field component $\phi$ has been introduced which couples with fluid 4-velocity $ U^{\mu} $ as $ U^{\mu} \p_{\mu} \phi $. In the previous works \cite{Chatterjee:2021ijw,Hussain:2022osn} the present authors introduced the topic of non-minimal coupling between the $k-$essence field and an ideal, relativistic hydrodynamic fluid. In those works the field-fluid interaction term was independent of the first derivative coupling $U^{\mu} \p_{\mu} \phi$. In this paper we see that inclusion of the first derivative coupling modifies the previous results and requires a new formulation. Due to the presence of  $U^{\mu} \p_{\mu} \phi$ factor, the field gradient directly couples with the fluid 4-velocity. Due to the presence of this coupling, the background field equations get modified. The derivative dependent coupling term $f(n,s,\phi) J^{\mu}\partial_{\mu} \phi$ was initially introduced in the context of quintessence field coupled with dark matter in  Ref.~\cite{Boehmer:2015sha, Koivisto:2015qua}. In our approach, we have extended the interaction functional form to  \(f(n,s, \phi, X)J^{\mu}\partial_{\mu} \phi\). Our ignorance about the proper phenomenology of the dark sector forces us to rely more on the formal understanding of the system. The derivative coupling is motivated by this formal urge. Except for the algebraic coupling, this is another form of non-minimal coupling which the dark sector can have in principle. Which one out of these two is physically realized can only be ascertained after more phenomenological results are obtained. The derivative coupling has its own particular properties which are distinct from the algebraic coupling. In the present model, the non-minimal coupling is suppressed when the scalar field attains an extrema or becomes constant in any phase of cosmological evolution. These properties are not present in the algebraic non-minimal coupling case. The resulting coupling of fluid velocity $U^{\mu}$ with the space-time derivative of field results in more complicated dynamics (when compared to the algebraic coupling) at the background and perturbation level. Any variations in \( U^{\mu}\) would directly propagate to the field $\phi$ and vice-versa. These kinds of models may play a relevant role in the theories related to linear growth of large-scale structures. It would be of particular interest to test models based on this type of interaction against recently published data \cite{Planck:2015bue, Planck:2018vyg}.  This interaction term also gives rise to nontrivial non-linear effects on the dynamics of the early phase of the universe.

Using the above mentioned field-fluid coupling  in our cosmological model we will present the dynamical stability analysis by advancing two different kinds of interacting models (corresponding to two different choices of $f$) in the presence of inverse square law $k$-essence scalar field potential $ V(\phi) \propto 1/\phi^2 $. In recent times the stability analysis approach in dark energy models has become useful \cite{Boehmer:2015sha, Yang:2010vv, Pal:2019tkc,Chakraborty:2019swx,Roy:2017uvr,DeSantiago:2012nk}. In the present work we will show the behavior of the non-minimally coupled model. It will be shown that the presence of non-minimal coupling can produce different phases of the universe. The coupled model can produce both phantom and non-phantom types of dark energy. 

The topics discussed in the paper are presented in the following manner. The basic model equations and analysis of the system have been presented in sec.~\ref{Theoretical Framework}. In sec.~\ref{Conservation sectors}  we show the energy-momentum conservation of the field-fluid system. The interacting field-fluid theory has been formulated in the context of a flat FLRW universe in sec.~\ref{Int FLRW Cosmology}. This coupled system's dynamical stability has been thoroughly investigated using two kinds of interacting models in
sec.~\ref{Cosmological Dynamics}. A generalized conclusion has been drawn in
sec.~\ref{Conclusion}.

\section{The basic formulation }
\label{Theoretical Framework}

The introduction of the $k$-essence theory does not only describe the early universe scenario; rather, this theory is essentially sound to explain the late-time cosmic behavior. Several articles \cite{ArmendarizPicon:1999rj, ArmendarizPicon:2000ah} have already encompassed the dual usefulness of this non-canonical scalar field theory. Main ingredients of this theory are a scalar field $\phi$ and a non-canonical kinetic term, $X=-\frac12 g^{\mu\nu} \nb_\mu \phi \nb_\nu \phi\, $. In this section we construct the interacting model from the Lagrangian approach where scalar field $\phi$ couples with a relativistic fluid non-minimally. Hence the total action of this coupled sector can be written as follows:
\begin{multline}
	S = \int_\Omega d^4x \bigg[\sqrt{-g}\frac{R}{2\kappa^2}-\sqrt{-g}\rho(n,s) +
	J^\mu(\varphi_{,\mu} + s\theta_{,\mu} + \beta_A\alpha^A_{,\mu})    
	-\sqrt{-g}{\mathcal L}(\phi,X)\\
	+ f(n,s,\phi,X)J^{\mu}\p_{\mu}\phi\bigg]\,.
	\label{act}
\end{multline}
Here the First term of this total action is the Einstein-Hilbert action, where $ R $ is the Ricci scalar, $ \kappa^2 = 8 \pi G $ and $ g $ specifies the determinant of metric $ g_{\m} $. We have considered the metric
$$ds^2=-dt^2 + a(t)^2 d{\bf x}^2\,,$$
with an isotropic and homogeneous (FLRW) background throughout the paper. Here $a(t)$ is the scale factor in the FLRW spacetime. The next two terms of the grand action represent the relativistic fluid introduced by Brown \cite{Brown:1992kc}. The action consists of energy density term, $\rho(n,s)$ which depends on particle number density, $n$ and entropy density per particle, $s$. Other parameters $\varphi, \theta, \alpha^A, \beta_A$ are all Lagrangian multipliers. The current density $ J^{\mu} $ is related with the particle density $n$ as, 
\begin{equation}\label{}
	J^\mu = \sqrt{-g}\, n\,U^\mu ,  \quad |J|=\sqrt{-g_{\mu\nu}J^\mu J^\nu}, \quad n=\dfrac{|J|}{\sqrt{-g}}, \quad U^\mu U_\mu=-1\,.
\end{equation}
where  $U^{\mu} = (1, \bm{0})$ specifies the 4-velocity of a fluid parcel. 
The action of $k$-essence scalar field is specified by the fourth term in the action and the final term signifies the coupling between the scalar field and the relativistic fluid. The coupling term $ f $ is an arbitrary function of $ (n,s,\phi, X) $ and it is accompanied by the first derivative of the field $\phi$, which is contracted with fluid current density $ J^{\mu} $. Because of this type of construction, the coupling term $ f $ has mass dimension one.  

To find the  equation of motion of the grand action we vary the Lagrangian in Eq.~(\ref{act}) with respect to the dynamical variables $g_{\m}, J^{\mu},s, \theta,\varphi, \beta_{A}, \alpha^{A}, \phi$. The variation of the action with respect to $ g^{\m} $ gives the energy momentum tensor  
\begin{eqnarray}
	T_{\mu \nu} \equiv \dfrac{-2}{\sqrt{-g}}\dfrac{\delta S}{\delta g^{\mu \nu}}\,.
	\label{energy momentum def}
\end{eqnarray}
Therefore the energy-momentum tensor of the relativistic fluid, $ k $-essence field and interaction becomes 
\begin{eqnarray}
	T^{\mu \nu}_{(M)} &= & 	 n\dfrac{\p \rho}{\p n} \left[U^{\mu} U^{\nu} + g^{\m} \right]-g^{\m} \rho\,,  \label{fluid energy momentum} \\
	T^{\mu \nu}_{(\phi)} &= &- {\mathcal L}_{,X}\,(\partial^\mu \phi)(\partial^\nu \phi)-g^{\mu \nu}\,{\mathcal L}\,, \label{tphi} \\
	T^{\m}_{(\rm int)} &= & \left[- n^2 \dfrac{\p f}{\p n}\left[U^{\mu} U^{\nu} + g^{\m} \right] +n \dfrac{\p f}{\p X} (\p^\mu \phi \p^\nu \phi)  \right] U^{\alpha}\p_{\alpha}\phi \,.  \label{inetraction energy momentum}
\end{eqnarray}
Comparing these energy momentum tensor with the perfect fluid $ T^{\m} = \rho U^\mu U^{\nu}+P (g^{\m} + U^{\mu} U^{\nu}) $, we get energy density and pressure of the fluid from \eqq{fluid energy momentum} 
\begin{equation}\label{}
	\rho_{M} = \rho, \quad 	P_{M} =  \left( n\dfrac{\p \rho}{\p n} - \rho \right) \,.
\end{equation}
The $ k $-essence energy density and pressure from \eqq{tphi} as 
\begin{equation}\label{}
	\rho_{\phi}= -{\mathcal L}_{,X}(\dot{\phi}^2) + \mathcal{L}\,, \quad P_{\phi} = - \mathcal{L}
\end{equation}
The interaction energy density and pressure can be written from Eq.~\eqref{inetraction energy momentum} as,
\begin{equation}\label{}
	\rho_{\rm int}  = n\dfrac{\p f}{\p X}\dot{\phi}^3 \,, \quad P_{\rm int} = \left[- {n^2} \dfrac{\p f}{\p n} \right]\dot{\phi} \,.
\end{equation}
These three individual sectors' energy-momentum tensors jointly produce the total energy-momentum tensor of the coupled sector, which can be presented as,
\begin{equation}\label{}
	T^{\mu \nu}= T^{\mu \nu}_{(M)} + T^{\mu \nu}_{(\phi)} + T^{\mu \nu}_{({\rm int})}\,.
\end{equation}
Varying the action in \eqq{act} with respect to fluid parameters gives
\begin{align}
	J^{\mu}:& \quad	\mu U_{\mu} + \left( \varphi_{,\mu}+ s\, \theta_{,\mu} + \beta_A \alpha_{,\mu }^A\right)  + \left[ \left(- n\dfrac{\p f}{\p n} \right) U^{\nu}U_{\mu} + f \delta _{\mu}^{\nu} \right] \p_{\nu}\phi = 0 \,,
	\label{jeqn}\\ 
	s:& \qquad -\dfrac{\p \rho}{\p s} + \dfrac{\p f}{\p s} n U^{\mu} \p_{\mu} \phi + nU^{\mu} \theta_{,\mu} = 0\, \label{entropy eqn} \,,\\  
	\varphi:& \qquad J^\mu{}_{,\mu}=0 \,,\label{number conservation}\\
	\theta:& \qquad (sJ^\mu)_{,\mu}=0 \,,\label{entrop conservation}\\
	\beta_A:& \qquad J^\mu\alpha^A_{,\mu}=0 \,,\label{fluid current along alpha}\\
	\alpha^A:& \qquad (\beta_AJ^\mu)_{,\mu}=0 \,,\label{fluid current conservation}
\end{align}
where $\mu$ signifies the chemical potential and defined as $\mu = \dfrac{\p \rho}{\p n} = \dfrac{\rho + P}{n}$. Because of the interaction term  the fluid dynamical equations  (\eqq{jeqn} and \eqq{entropy eqn}) become generalized. Eq.~\eqref{number conservation}
shows that the particle number of the system is conserved, whereas,  Eq.~\eqref{entrop conservation} indicates that there is no flow of entropy across the fluid elements. Hence the system is in an adiabatic process. Together they constrain the system and we can write
\begin{equation}
\nb_{\mu }(n U^{\mu}) = 0 \quad {\rm and} \quad  \nb_{\mu}(n s U^{\mu}) = 0\,,
\label{eq:constraint relation}
\end{equation}
where $\nb_{\mu }$ is covariant derivative. Suppose in the action the interaction term does not exists i.e., $ f=0 $. Hence from the first law of thermodynamics one can define the temperature of the fluid as $ T = \dfrac{1}{n} \left( \dfrac{\p \rho}{\p  s }\right) $ \cite{Brown:1992kc}. In that case Eq.~\eqref{entropy eqn} gives $ U^\mu \theta_{,\mu} = \dfrac{1}{n} \left( \dfrac{\p \rho}{\p  s }\right) = T $, hence, $\theta$ behaves as the potential for temperature. Using Eq.~\eqref{jeqn}, \eqq{entropy eqn} and \eqq{fluid current along alpha} one can show that the $ F = \mu  - Ts  = \varphi_{,\mu} U^\mu $, where $ F $ is the chemical free energy and $\varphi$ signifies the potential of chemical free energy.\footnote{ One shouldn't confuse with chemical free energy $ F $ with the $ k $-essence kinetic term $ F(X) $, to be introduced later, they are completely different quantities.} Because of the presence of interaction in the system i.e., $ f \ne 0 $,  \eqq{entropy eqn} gives 
\begin{equation}\label{}
U^\mu \theta_{,\mu} = \dfrac{1}{n} \left( \dfrac{\p \rho}{\p  s } - \dfrac{\p f }{\p s} n U^{\mu} \p_{\mu}\phi \right) = T + T_{\rm int}\,. 
\end{equation}
Using Eq.~\eqref{jeqn}, \eqq{entropy eqn} and \eqq{fluid current along alpha} the potential for chemical free energy $ \varphi $ gets generalized
\begin{equation}\label{}
	\varphi_{,\mu} U^\mu  = \mu - sT + U^{\alpha} \p_{\alpha} \phi \left(s \dfrac{\p f}{\p s} - n \dfrac{\p f}{\p n} -f \right) = F + F_{\rm int}\,.
\end{equation}
The above equations specify how the non-minimal field-fluid coupling affects the basic fluid equations.

Varying the action in Eq.~\eqref{act} with respect to $\phi$ produces the modified field equation. Due to the presence of non-minimal interaction term the field equation becomes:
\begin{multline}
	\bigg[- \dfrac{\p \mathcal{L}}{\p \phi} +  \dfrac{\p f}{\p \phi}  \, n U^{\alpha}\p_{\alpha}\phi - \p_{\mu} \left( \dfrac{\p \mathcal{L}}{\p X} \p^\mu \phi \right)  + \p_{\mu}\left( \dfrac{\p f}{\p X}(\p^\mu \phi) \right) (n U^{\alpha} \p_{\alpha} \phi) \\  + f_{,X} (\p^{\mu}\phi ) \p_{\mu}(nU^\alpha \p_{\alpha}\phi )  - \p_{\mu}(f) n U^{\mu} 
	\bigg] = 0\,.
\end{multline} 
This equation can be expressed in terms of covariant derivative as: 
\begin{multline}\label{field equation of kessence}
	\bigg[ - \dfrac{\p \mathcal{L}}{\p \phi}  - \nb_{\mu} \left( \dfrac{\p \mathcal{L}}{\p X} \p^\mu \phi \right) 
	+ \dfrac{\p f}{\p \phi}  \, n U^{\alpha}\p_{\alpha}\phi 
	+ \nb_{\mu}\left( \dfrac{\p f}{\p X}(\p^\mu \phi) \right) (n U^{\alpha} \p_{\alpha}\phi)
	\\
	+ f_{,X}(\p^\mu \phi) \nb_{\mu}(nU^{\alpha}\p_{\alpha}\phi) 
	\bigr\}- (\nb_{\mu}f)   (nU^{\mu})    
	\bigg] = 0 \,. 
\end{multline}
Before we end this section we want to specify that in this paper heat exchange between various fluid parcels will be assumed to be zero. The whole fluid is assumed to be having the same temperature. The constraints in Eq.~(\ref{eq:constraint relation}) specify that $\dot{s}=0$.  Henceforth we will always assume that the entropy density of the fluid is constant.
\section{Conservation of energy-momentum tensor in presence of non-minimal coupling}
\label{Conservation sectors}

Due to the presence of the non-minimal coupling term, the individual energy-momentum tensors, in general, are not conserved, but the total energy-momentum tensor is conserved, hence $ \nb_{\mu}T^{\m}=0 $. To prove the conservation of total energy-momentum tensor we regroup the various energy-momentum tensors and define two new energy-momentum tensors as: 
$ T^{'\m} = T_{(\phi)}^{\m} +  n \dfrac{\p f}{\p X} (\p^\mu \phi \p^\nu \phi)   U^{\alpha}\p_{\alpha}\phi$ and $\tilde{T}^{\mu \nu} = T_{(M)}^{\m} + T_{({\rm int})}^{\m} - n \dfrac{\p f}{\p X} (\p^\mu \phi \p^\nu \phi)   U^{\alpha}\p_{\alpha}\phi$. 
Using these redefined tensors we can now write the Einstein field equation as:
\begin{equation}\label{}
G^{\m} =\kappa^2 (T'^{\mu\nu}+\tilde{T}^{\mu\nu})\,.
\end{equation}
Now we can show that $ \nb_{\mu}T^{'\m} =Q^{\nu} $ and $ \nb_{\mu}\ti{T}^{\m} = - Q^{\nu} $. To show that we will first calculate $\nabla_{\mu}T^{'\m}$.  Using Eq.~\eqref{field equation of kessence} we get  
\begin{eqnarray}
\nabla_{\mu}T^{'\m} &=&   	\bigg[f_{,n}(\nb_{\mu}n) + f_{,X}(\nb_{\mu}X)  \bigg] (nU^{\mu} )\p^{\nu}\phi - f_{,X}(\nb^{\nu}X)  (nU^{\alpha} \p_{\alpha}\phi) =  Q^{\nu}\,.
\label{Conserved scalar field}
\end{eqnarray}
Now to show $ \nb_{\mu}\ti{T}^{\m} = - Q^{\nu} $, we express the covariant derivative into parallel and perpendicular components of the fluid flow as:
\begin{equation}\label{Fluid emt def}
\nb_{\mu}	\ti{T}^{\m} = h^{\nu}_{\lambda}\nb_{\mu} \ti{T}^{\mu \lambda} - U^\nu U_\lambda \nb_{\mu}\ti{T}^{\mu \lambda}\,, 
\end{equation}
where $h_{\m} = g_{\m} + U_{\mu}U_{\nu }$. Using $ \nb_{\mu}(nU^{\mu}) = 0 , \; \dot{s} = 0$, $ U_{\nu}\nb_{\mu}U^{\nu} = 0  = U^{\nu}\nb_{\mu}U_{\nu}$ and $ U^{\nu}U_{\nu} =-1  $, one can show that the last term  of Eq.~\eqref{Fluid emt def} can be written as
\begin{equation}\label{perpendicular component}
U_\nu \nb_{\mu}\ti{T}^{\,\m} 	 =  n^2\dfrac{\p f}{\p n} U^\alpha \p_{\alpha}\phi \nb_{\mu}U^\mu\,.
\end{equation}
One can also rewrite the first term of Eq.~\eqref{Fluid emt def} in a different way. The manipulations involving the first term on the right hand side (of Eq.~(\ref{appeqn})) are explicitly shown in appendix \ref{proof01}. Using the result obtained in the appendix we can write
\begin{eqnarray}\label{appeqn}
\begin{split}
\nb_{\lambda}\ti{T}^{\beta \lambda}  = h^{\beta}_{\nu}\nb_{\lambda}\ti{T}^{\nu \lambda} - U^{\beta}U_{\nu}\nb_{\lambda}\ti{T}^{\nu \lambda} 
= g^{\beta \mu}h_{\m}\nb_{\lambda}\ti{T}^{\nu \lambda}  - U^\beta n^2\dfrac{\p f}{\p n} U^\alpha \p_{\alpha}\phi \nb_{\nu}U^\nu\\
= g^{\beta \mu} \big[\nb_\lambda  (nf_{,n} U^\nu \p_\nu \phi)n U^\lambda U_\mu + n U^\lambda (\nb_\lambda U_\mu )nf_n U^\nu \p_\nu \phi  -  \bigl\{ f_{,n} \nb_\lambda n\\
+ f_{,X} \nb_{\lambda} X \bigr\}  nU^\lambda \p_\mu \phi 
+n \nb_\mu  (nf_{,n} U^\nu \p_\nu \phi) + \left\lbrace f_{,n} \nb_\mu n + f_{,X} \nb_\mu X  \right\rbrace  nU^\lambda \p_\lambda \phi \big] \\
- g^{\beta \mu}   n^2f_{,n} U^\alpha \p_{\alpha}\phi (U^\lambda \nb_{\lambda}U_\mu) 
- g^{\beta \mu}   h^\lambda _\mu \nb_\lambda \left(n^2f_n U^\alpha \p_{\alpha}\phi \right) 
- U^\beta n^2\dfrac{\p f}{\p n} U^\alpha \p_{\alpha}\phi \nb_{\nu}U^\nu\,.               
\end{split}	
\end{eqnarray}
After some rearrangement one can show that 
\begin{equation}\label{final fluid conservation}
\nb_{\lambda}\ti{T}^{\beta \lambda}= Q^{\beta}=- \left( f_{,n} (\nb_\lambda n ) + f_{,X} (\nb_{\lambda} X) \right)  nU^\lambda \p^\beta \phi 
+ f_{,X} \nb^\beta X   (nU^\lambda )\p_\lambda \phi\,. 
\end{equation}
Comparing Eq.~\eqref{Conserved scalar field} and Eq.~\eqref{final fluid conservation} it can be seen that the total energy-momentum tensor is conserved as:
\begin{equation}\label{}
\nb_\mu (T'^\m +\ti{T}^\m) = 0\,.
\end{equation}

\section{The dynamical equations}
\label{Int FLRW Cosmology}

Using the above general framework of coupled field-fluid scenarios we investigate this system in the context of a spatially flat FLRW universe. The Friedmann equations get modified due to the presence of an interacting scenario. The Friedmann equation and other equations specifying the fluid, like the equations specifying the conservation of particle number density and entropy per particle, can be written as:
\begin{eqnarray}
3H^2 &=& \kappa^2 \left(\rho+\rho_{\phi}+\rho_{\rm int}\right) = \kappa^2 \rho_{\rm tot}\,,
\label{frd1}\\
2\dot{H}+3H^2 &=& -\kappa^2 \left(P+P_{\phi}+P_{\rm int}\right) = -\kappa^2 P_{\rm tot}\,.
\label{eq:F1} \\
\dot{n}+3Hn &= & 0,\quad {\rm and} \quad \dot{s}=0\,.
\end{eqnarray}
Here $H=\dot{a}/a$ is the Hubble parameter. We have used the form of $k$-essence Lagrangian, ${\mathcal L}=-V(\phi)F(X)$ where $F(X)$ is a non-canonical kinetic function of $X$ and potential $V(\phi)$ depends on the scalar field $\phi$. Hence the energy density and pressure of the $k$-essence sector can be written as
\begin{eqnarray}
\rho_{\phi} &=& V(\phi)(2XF_{,X}-F)\,,\quad P_{\phi}=V(\phi)F(X)\,.
\label{eq:F2}
\end{eqnarray}
The equation of state (EOS),  $\omega_\phi$, and adiabatic sound speed $c_{s(\phi)}^2$ in the $k$-essence sector are:
\begin{eqnarray}
\omega_\phi = \frac{P_{\phi}}{\rho_{\phi}}\,,\quad c_{s(\phi)}^2 = \frac{dP_\phi/dX}{d\rho_\phi/dX}\,.
\label{eq:F3}
\end{eqnarray}
In the absence of an interacting term the above EOS determines the nature of the system at late times. The sound speed limit put additional constrained on the system as $ 0 \le c_{s(\phi)}^2 \le 1 $. For negative sound speed the system becomes unstable.

The modified scalar field equation, Eq.~\eqref{field equation of kessence}, can be expressed as:
\begin{multline}
	\label{field kessence}
	-\mathcal{L}_{,\phi} + \mathcal{L}_{,\phi X}\dot{\phi}^2 + \mathcal{L}_{,XX}\dot{\phi}^2 \ddot{\phi} + 3H\mathcal{L}_{,X} \dot{\phi} + \mathcal{L}_{,X}\ddot{\phi} + 
	\bigg[3nH f_{,nX}\dot{\phi} - f_{,\phi X} \dot{\phi}^2 \\ - f_{,XX} \dot{\phi}^2 \ddot{\phi} - f_{,X} \ddot{\phi} \bigg]\, n \dot{\phi} - 
	n f_{,X} \dot{\phi} \ddot{\phi} + 3n^2 H f_{,n}  - n f_{,X} \dot{X} = 0\,.
\end{multline}
where subscript $XX$ or $\phi X$ implies a second-order partial derivative with respect to $X$ or $\phi$ and $X$. The $ Q^{\nu} $ in this background metric will become $  Q^{0} = n^2f_{,n} 3 H \dot{\phi} $. Therefore the field-fluid exchanges energy density via this coupling. In the next section, we will set up the dynamical approach to study the coupling scenario. 
%
%
\section{Dynamical stability of non-minimally coupled system}
\label{Cosmological Dynamics}

In this section we will construct the dimensionless dynamical variables to study the dynamics of the interacting system. The basic dynamical variables are $ x,y,z,\sigma $ which are defined as
\begin{equation}
x = \dot{\phi}, \quad y = \frac{\kappa\sqrt{V(\phi)}}{\sqrt{3}H}, \quad \sigma = \frac{\kappa\sqrt{\rho}}{\sqrt{3}H}, \quad z = \frac{\kappa^2 nf}{3H^2}
	\label{eq:D0}
\end{equation}
The other variables $A, B, C, D, E $ can be expressed in terms of the basic variables as: 
\begin{eqnarray}
	\quad B = \frac{nf_{,\phi}\kappa^2}{H^3}, \quad E= \dfrac{n\kappa^2}{H^3} \frac{\partial^2 f}{\partial \phi \partial X} ,\quad D =  \dfrac{ n\kappa^2 }{3H^2}f_{,X}, \quad A =  -\dfrac{n^2\kappa^2}{H^2} \frac{\partial f}{\partial n},
	\quad C= \dfrac{\kappa^2 P_{\rm int}}{3H^2}
	\label{eq:D1}
\end{eqnarray}
Due to the presence of a constraint equation only $ x,y,z$ are used to study the dynamical system. 
The constrained
equation of the system can be written by using Eq.~\eqref{frd1} as:
\begin{eqnarray}
\sigma^2 =  1- y^2\left(3x^4/4-x^2/2\right) -  x^3 D\,.
\label{eq:dyn1}
\end{eqnarray}
where $\sigma$ denotes the fluid variable and it satisfies the inequality $ 0 \le \sigma^2 \le 1 $. The other Friedmann equation can be expressed as 
\begin{eqnarray}
\dfrac{\dot{H}}{H^2} = -\frac32 [\omega \sigma^2 + y^2 F+C+1]\,.	\label{eq:dyn2}		
\end{eqnarray}
The dynamical equations can be expressed compactly if we represent the dimensionless energy density of the scalar field as $\Omega_{\phi}$ and interaction energy density as $ \Omega_{\rm int} $
\begin{equation}\label{diminsionless_energy_den}
	\Omega_{\phi} = \frac{\kappa^2 \rho_{\phi}}{3H^2} = y^2 (x^2 F_{,X} - F)\,, \quad  \Omega_{\rm int} = \frac{\kappa^2 \rho_{\rm int}}{3H^2} = D x^3\,.
\end{equation}
Using the variables defined above the autonomous equations of the system are: 
\begin{subequations}\label{generalized autonomous eqn}
\begin{eqnarray}
x^{\prime}=	\frac{\dot{x}}{H}  &= & \dfrac{\lambda F- \lambda F_{,X}x^2-3 x y^2F_{,X}-x^2A_{,X}-\frac{E}{3} x^3 - A}
{[y^2F_{,XX}+ D_{,X} x]x^2 + [y^2 F_{,X} + 3 D x]}\label{auto_x}\\
y^{\prime} = \frac{\dot{y}}{H}  &=& \dfrac{\lambda x }{2y} + \dfrac{3}{2}y  \left(\omega \sigma^2 + y^2 F+C+1 \right)\label{auto_y}\\
z^{\prime} = \frac{\dot{z}}{H}  &=&3z\left(\omega \sigma^2 + y^2 F+C \right) + A + \frac{Bx}{3} + D x x^{\prime}\label{auto_z}
\end{eqnarray}
\end{subequations}
The system possesses 3-D autonomous equations for any general interaction. 
Here prime signifies the derivative of the dynamical variables $x,y,z$ with respect to $H dt$. 
In these equations $\lambda$ is the dimensionless parameter which signifies the derivative of $k$-essence sector potential  $\lambda = \dfrac{\kappa^2 V_{,\phi}}{3H^3} $ .
Corresponds to these autonomous equations the critical points $ (x_{0}, y_{0}, z_{0}) $ can be found by solving $x^\prime =0$, $y^\prime=0$ and $z^\prime=0$. 
To understand the behavior of the system near these critical points we linearize the autonomous equation around these critical points using the Taylor series expansion up to the first order and hence find the Jacobian matrix.
Based on the eigenvalues of the matrix the nature of fixed points can be established. The $ 3 \times 3 $ matrix posses three eigenvalues and if all the eigenvalues have negative (positive) real parts then the fixed point is asymptotically stable (unstable). If any of these eigenvalues which have a positive real part and the rest have a negative real part; the fixed points become saddle. If any of the eigenvalues turns out to be zero, the linear stability theory fails, and one needs to employ the centre manifold theory to further understand the nature of those critical points. 

To further study the autonomous equations we need to introduce the form of interaction and potential. Hence we will  use a power-law type potential of $k$-essence which has been studied extensively \cite{ArmendarizPicon:2000ah, Chiba:1999ka, Chiba:2002mw, Kang:2007vs} as:
\begin{eqnarray}
V(\phi) = \dfrac{\delta}{\kappa^2\phi^2}\,.
\label{plpot}
\end{eqnarray}
Here $\delta$ is a dimensionless parameter. This type of inverse square law scalar field-dependent potential is widely used to probe the late time accelerating universe in  cosmological models employing $k-$essence theory. We have also introduced two different interaction models in Tab.[\ref{table:M1}]. The form of $f$ as tabulated are some of the simplest choices one can make in the present context. Based on these forms of interaction we will establish the dynamics in the next section. To proceed further we have chosen the form of the kinetic part of the $k$-essence field as: 
\begin{equation}
F(X) = -X+  X^2\,.
\label{Fform}
\end{equation}
This form of $F$ is also widely used in the literature.
\begin{table}[t]
	\centering
	\begin{tabular}{|p{0.5cm}| p{1.9cm} |p{2.5cm}|p{0.8cm}|p{1.2cm}|p{1cm}|p{1cm}|p{3cm}|p{1.5cm}|}
		\hline 
		\multicolumn{9}{|c|}{ $V(\phi) = \dfrac{\delta}{ \kappa^2 \phi^2} $  } \\
		\hline
		SR &	$ f $  & $ B $ &$  D $ & $D_X$ & $ C $ & $ A $ & $ E  $ &  $\lambda$ \\
		\hline
		&&&&&&&& \\
		I &	$\frac{\rho}{n} (\phi /\kappa )^m X^{\eta} $ & $- 3\sqrt{3} z\, m  \dfrac{ y}{\delta^{1/2}}  $ &  $ \dfrac{2{\eta}\,z}{x^2} $ & $\frac{4{\eta}({\eta}-1)z}{x^4} $ & $- \omega z x$ &$ -3z \omega$ & $- \dfrac{ 6 m \eta z }{x^2} \left( \dfrac{3^{1/2} y}{\delta^{1/2}} \right) $ & $\dfrac{- 2 \sqrt{3} y^{3}}{ \delta^{1/2}} $ \\
		
		\hline
		II. &$  \dfrac{F}{n \kappa^2 \phi^2} $ & $ \dfrac{-6 y^3 \sqrt{3} F}{\delta \sqrt{\delta}} $ & $ \dfrac{y^2 F_{,X}}{\delta} $ & $ \dfrac{y^2 F_{,XX}}{\delta} $ & $ \dfrac{y^2 F x}{\delta} $ & $ \dfrac{3 y^2 F}{\delta} $ & $ \dfrac{-6\sqrt{3} y^3 F_{,X}}{\delta \sqrt{\delta}} $ & $\dfrac{- 2 \sqrt{3} y^{3}}{ \delta^{1/2}} $ \\
		
		\hline
	\end{tabular}
	\caption{Models parameters of the interacting $ k $-essence field-fluid. }
	\label{table:M1}
\end{table}

\subsection{Study of Model I}
To study the field-fluid coupling of the system we have chosen the form of interaction $ f= \dfrac{\rho}{n} (\phi /\kappa )^m X^{\eta} $, where $ \rho $ signifies the matter-energy density, $ n $ is the fluid number density and $\phi$, $ X $ depend on the field and its time derivatives. Here $ m $ and $\eta$ are the model parameters and can take any value based on the system dynamics. The total energy density and pressure for this system can be expressed as:
\begin{eqnarray}
	\rho_{\rm tot} &=& \rho + \rho_{\phi} +\rho_{\rm int} =\rho + V(\phi) (2XF_{,X}-F) +n\frac{\p f}{\p X}\dot{\phi}^3\,,  \label{rho_tot_mod1} \\
	P_{\rm tot} &=& P + P_{\phi} + P_{\rm int} = P + V(\phi)F   -\left( n^2 \frac{\p f}{\p n} \right)\dot{\phi}\,. 
\end{eqnarray}
Constraint equation for this model can be written as:
\begin{eqnarray}
	\sigma^2 =  1- y^2\left(\frac34 x^4-\frac12 x^2\right) -   2x\eta z\,.
	\label{eq:dynm1}
\end{eqnarray}
Using the above expressions we can define the grand EOS and adiabatic sound speed as:
\begin{eqnarray}
	\omega_{\rm tot}	&=& \frac{P_{\rm tot}}{\rho_{\rm tot}} = \omega \sigma^2 + y^2\left(\frac14 x^4-\frac12 x^2\right) +C \,,\label{eos}\\
	c_s^2 &=& \dfrac{dP_{\rm tot}/dX}{d\rho_{\rm tot}/dX} = \dfrac{y^2 F_{,X} - 2 \omega z(\eta + 1/2)/x}{y^2(F_{,X} + 2XF_{,XX}) + \frac{4\eta (\eta -1)z }{x} + \frac{6 \eta z}{x}}\,. 
	\label{eq:M1-d}
\end{eqnarray}

\subsection{Dynamics of the Model}

We will study this model in the presence of both radiation ($ \omega = 1/3 $) and matter ($ \omega = 0 $). The 3-D autonomous system is complicated and contains three model-dependent parameters $\delta, m$ and $\eta$. Hence, finding the critical points in model-dependent parameters is a rather complicated task. To resolve this issue, we will use the following technique. We have checked that our system contains eight critical points, when we deal with dust and six critical points, when we deal with radiation. Some of these critical points does not explicitly depend upon $m$ and $\eta$. The others may depend upon all the three parameters. We  figure out the critical points which are independent of specific values of  $m$ and $\eta$ in a particular way and later find out the other critical points in a more conventional approach.

To find out the $m$ and $\eta$ independent critical points we have expressed $\omega_{\rm tot}$, at a critical point, in terms of $ \gamma $, which is an arbitrary constant. We have replaced (the functional form of the kinetic term of $ k $-essence scalar field) $ F $, at the critical point, in terms of other variables present in the expression of Eq.~\eqref{F replacement}. We have then
\begin{eqnarray}
\omega_{\rm tot} & = & -1 + \gamma, \label{paraomega}\\
F & = &\dfrac{-1 + \gamma - \omega \sigma^2 - C}{y^2}\,.\label{F replacement}
\end{eqnarray}
Using these relations, at the critical points, we find out some of the critical points which do not explicitly depend on the values of $m$ and $\eta$.

\subsubsection{Dynamics in presence of dust}

\begin{figure}[t]
	\centering
	\includegraphics[scale=0.8]{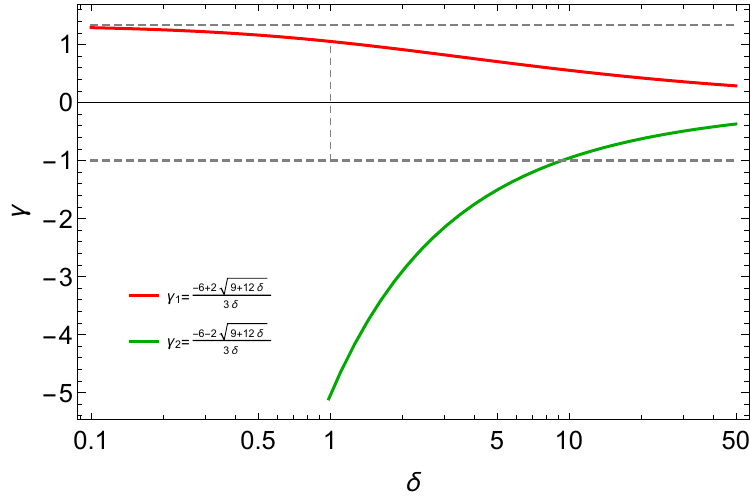}
	\caption{Variation of $\gamma$ against $ \delta $ for Model I.  }
	\label{fig:gamma delta}
\end{figure}

By replacing $ F $ in Eqs.~\eqref{auto_x}-\eqref{auto_z} by the above
value we get one set of the critical point $ P_{\mp} $ of the system
which is solely dependent on $ \gamma $ and $\delta$.  The general
form of the ($m$ and $\eta$ independent) critical points in the
matter-dominated background have been tabulated in the top row of
Tab.~[\ref{tab:temp critical points model 01}].  Using this set of
critical point $ P_{\mp} $ in the expression of the EOS, in Eq.~\eqref{eos}, we find the relation between $\gamma$ and $\delta$ shown in Fig.[\ref{fig:gamma delta}].
There exist two solutions of $\gamma$ as $\gamma_{1}$ and
$\gamma_{2}$ which characterize the non-phantom and phantom
solutions, respectively, for one set of $\delta$ values.  Inserting the value
of $ \gamma_{1} $ and $\gamma_{2}$ in the critical point expression $ P_{\mp} $
produces the four actual $\delta$ dependent critical points listed in Tab.[\ref{tab:temp
critical points model 01}], which are
labeled as $ P_{1\mp} $ and $ P_{2\mp} $.
%
\begin{table}[t]
	\begin{center}
		\centering
		\begin{tabular}{|c|c|c|c|}
			\hline
			\makecell{Critical \\ Points} & $ (x,y,z) $ & $ \omega_{\rm tot} $ & $ \gamma $\\
			\hline
			$ P_{\mp} $ & $ \left( \mp\dfrac{\sqrt{3 \gamma  \delta +4}}{\sqrt{3} \sqrt{\gamma } \sqrt{\delta }}, \mp \dfrac{3 \gamma ^{3/2} \delta }{2 \sqrt{3 \gamma  \delta +4}}, 0 \right)  $ & $ \frac{1}{16} \gamma  (4-3 \gamma  \delta ) $ & $  \dfrac{-6 \mp 2 \sqrt{12 \delta +9}}{3 \delta }$\\
			\hline
			$ P_{1\mp} $ & $ \left(\mp \frac{\sqrt{-\sqrt{12 \delta +9}-1}}{\sqrt{\delta } \sqrt{-\frac{\sqrt{12 \delta +9}+3}{\delta }}}, \mp \frac{\delta  \left(-\frac{\sqrt{12 \delta +9}+3}{\delta }\right)^{3/2}}{\sqrt{3} \sqrt{-\sqrt{12 \delta +9}-1}},0 \right)  $ & $ -\frac{3 \delta +2 \sqrt{12 \delta +9}+6}{3 \delta } $ & $ -\frac{2 \left(\sqrt{12 \delta +9}+3\right)}{3 \delta } $ \\
			\hline 
			$ P_{2 \mp} $ & $ \left(\mp\frac{\sqrt{\sqrt{12 \delta +9}-1}}{\sqrt{\delta } \sqrt{\frac{\sqrt{12 \delta +9}-3}{\delta }}}, \mp \frac{\delta  \left(\frac{\sqrt{12 \delta +9}-3}{\delta }\right)^{3/2}}{\sqrt{3 \sqrt{12 \delta +9}-3}},0 \right)  $ &$ \frac{-3 \delta +2 \sqrt{12 \delta +9}-6}{3 \delta } $ & $ \frac{2 \sqrt{12 \delta +9}-6}{3 \delta } $ \\
			\hline
			
		\end{tabular}
		\caption{Critical points in the matter dominated background for general $ m $ and $\eta$ for Model I. }
		\label{tab:temp critical points model 01}
	\end{center}
\end{table}

From Fig.[\ref{fig:gamma delta}] we see that for $0<\delta<1$, grand
EOS parameter shows a non-accelerating behavior corresponding to
critical point $ P_{2\mp} $, while $ P_{1\mp} $ shows 
phantom behavior. As $\delta$ increases, value of grand EOS parameter
corresponding to $ P_{2\mp } $ is going towards an accelerating EOS
($\gamma_{1} \to 0$) parameter while $ P_{1\mp } $ also converges
towards $ (\gamma_{2} \to 0)$ and produces less phantom behavior.
Hence, by choosing a larger value of $\delta$  one can find both
accelerating and phantom solutions.  Using this technique, we have
found four critical points which are independent of $m$ and $\eta$ for a large value of $\delta$. As the system has eight critical points we have to find four more critical points. It turns out that the other four critical points, called $ Q_{\mp} $ and $ R_{\mp} $, have to be found out from the equations $x^\prime=0$, $y^\prime=0$ and $z^\prime=0$ by. This critical points may depend on $m$ and $\eta$.
We have evaluated all the other critical points by choosing the benchmark
value of $\delta = 20$. For $\delta= 20$ the 
other sets of critical points $ Q_{\mp} $ and $ R_{\mp} $ are shown in
Tab.[\ref{tab:Model 01 matter critical points}], in which only $
Q_{\mp} $ are both $ m $ and $\eta$ dependent.  The other critical points $ R_{\mp} $ are evaluated numerically. For $ Q_{\mp} $, we
evaluated the grand EOS parameter, which turns out to be only $ m $
dependent. Consequently this critical point can be set to any phase of
the universe based on the value of $ m $. We have chosen to set
it in the radiation phase and $ m $ turns out to be $ m = 1/2 $. To
further constraint the model parameter $ \eta $ corresponding to $ m=1/2
$ we have used the sound limit, which puts a bound on $\eta$, and get $\eta \leq
{31}/{20}$. We have listed the all critical points corresponding
to the case where $\omega=0$ in Tab.~[\ref{tab:Model 01 matter
    critical points}] for $ \delta=20, m= 1/2 $ and $\eta = 1$ case.

\begin{figure}[t]
	\begin{minipage}[b]{0.5\linewidth}
		\centering
		\centering
		\includegraphics[scale=.5]{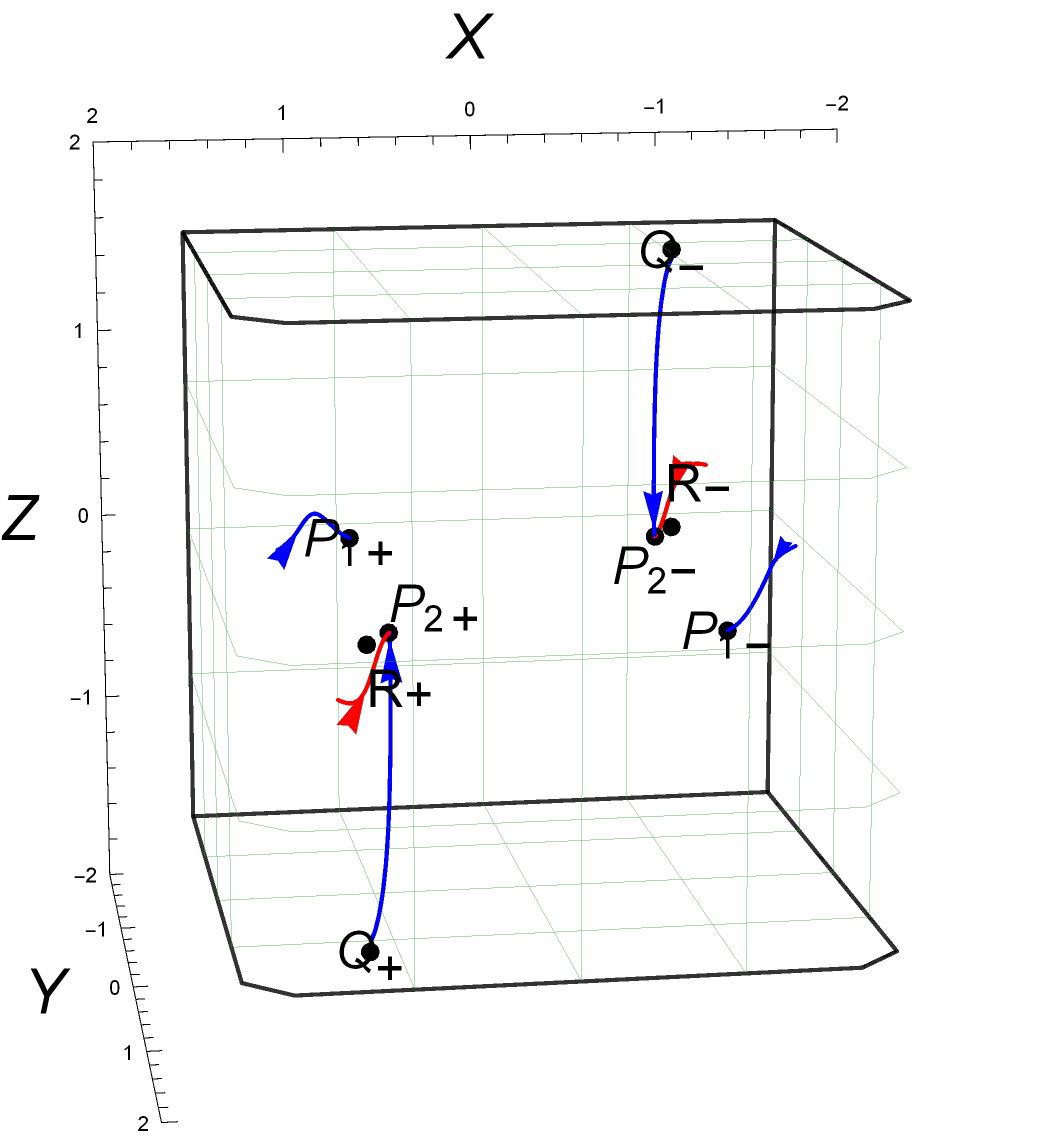}
		\subcaption{3D-Phase space for matter dominated fluid with $\delta = 20, m=1/2, \eta = 1$.}
		\label{fig:Model 01 matter phase space}
	\end{minipage}
	\hspace{0.2cm}
	\begin{minipage}[b]{0.5\linewidth}
		\centering
		\includegraphics[scale=0.65]{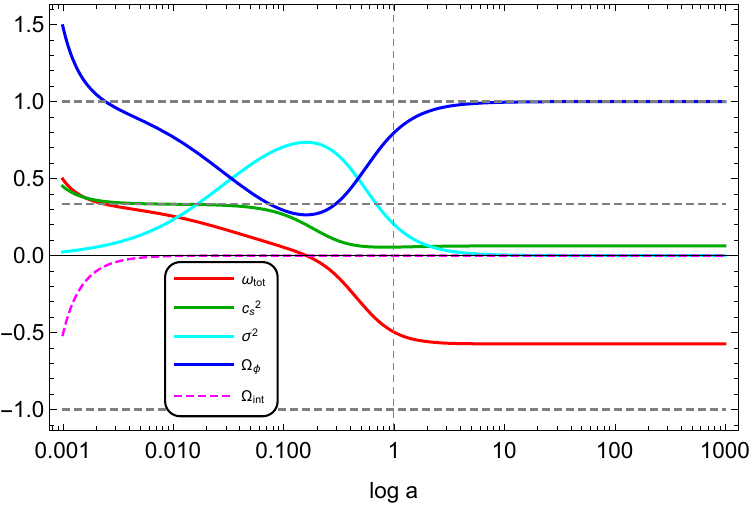}
		\subcaption{Evolution in the presence of matter background fluid with $\delta=20, m = 1/2$ and $ \eta = 1 $.  }
		\label{fig:Model 01 matter evolution}
	\end{minipage}
	\caption{3D Phase Space and Evolution plot in presence of dust for Model I.}
\end{figure}

We have verified that none of the critical points corresponding to $
w=0 $ exists at infinity. To understand the nature of the system, we
have portrayed the 3-D phase space and evolution of system parameters
with respect to the logarithmic function of the FLRW scale factor.  In
presence of dust two sets of critical points $P_{1\mp}$ and $P_{2\mp}$
are stable attractors.  Near one point the universe shows accelerating
phantom behavior with negative sound speed whereas near the other set
we have accelerating behavior as $ -1< \omega_{\rm tot}<-1/3 $ and the
sound speed is positive around the second point. The dynamics of the
critical points can be understood from the 3-D phase space plot in
Fig.~[\ref{fig:Model 01 matter phase space}] in which trajectories
corresponding to blue color show that they are attracted towards a
critical point while red-colored trajectories signify that they are
repelled from the critical points.

In the present case the phase space is not compact as $-\infty < x < \infty$, 
$-\infty < y < \infty$ and $-\infty < z < \infty$. To represent the phase space in a bounded region we have used a transformation of phase space variables. These transformations connect the conventional phase space variables $(x,y,z)$ to the new set of variables $(X,Y,Z)$ where the new variables are defined as:
\begin{eqnarray}
X=\tan^{-1}x\,,\,\,\,\,Y=\tan^{-1}y\,,\,\,\,\,Z=\tan^{-1}z\,.
\label{phasedef}  
\end{eqnarray}
These new variables map the unbounded phase space to a bounded region as
$$-\frac{\pi}{2} < X < \frac{\pi}{2}\,,\,\,\,\,-\frac{\pi}{2} < Y < \frac{\pi}{2}\,,\,\,\,\,-\frac{\pi}{2} < Z < \frac{\pi}{2}\,.$$
We use these new variables to redefine the phase space in both the cases, in presence of dust or radiation, in Model I.

We see that the trajectories nearby $ Q_{\mp} $ evolve towards
$P_{2\mp}$, showing that $ Q_{\mp} $ signifies saddle type critical
points. We can identify the points $ Q_{\mp} $ correspond to the
radiation phase of the universe. Hence all the nearby trajectories are
globally attracted towards $ P_{2\mp} $. The other trajectories nearby
the points $ R_{\mp} $, which signify the matter phase, are repelled
from that fixed points and eventually are attracted towards $ P_{2
  \mp} $. Few trajectories are attracted towards $ P_{1\mp} $, making
the point a stable attractor. Hence we see that for the dust solutions
the field-fluid scenario can successfully depict all four phases of
the universe.

We have evaluated the evolutionary dynamics of system parameters such as grand EOS ($\omega_{\rm tot}$), $ k $-essence energy density $ \Omega_{\phi} $, fluid density $ \sigma^2 $, sound speed $ c_{s}^2 $ and interaction density $ \Omega_{\rm int} $ and have shown their development in Fig.~[\ref{fig:Model 01 matter evolution}]. 
From the figure, we analyze that as the system enters the very early epoch the coupling between field-fluid becomes very strong and $ k $-essence density and interaction density start with a high magnitude. Since $ k $-essence energy density always has to be positive $ \Omega_{\phi} \ge 0 $, and system follows the constrained relation Eq.~\eqref{eq:dyn1}, hence $ \Omega_{\rm int} $ becomes negative. Due to the high energy density of $ k $-essence field, the grand EOS of the system tends to be stiff. The sound speed is also greater high initially. As the system evolves, the total EOS of the system becomes $1/3$, and the $ k $-essence energy density dominates over the fluid energy density $ \sigma^2 $ and $ \Omega_{\rm int} $. This phase is  representative of the radiation dominated phase. The fluid energy density, $ \sigma^2 $, then starts increasing and $ \Omega_{\phi} $ decreases and becomes subdominant in the matter phase; at this phase, the EOS becomes zero. The $ k $-essence energy density is not zero during the matter phase and hence keeps increasing and becomes dominant at the late-time phase. With this phase transition the grand EOS also decreases towards a negative value and gets saturated at $ \approx -0.6 $. The sound speed also decreases and becomes $ \approx 10^{-3} $. Hence the system transforms from stiff matter to the late-time cosmic acceleration when the hydrodynamic fluid represents dust. 

\begin{table}[t]
	
	\centering
	\begin{tabular}{|c|c|c|c|c|}
		\hline
		\makecell{Critical \\ Points} & $ (x,y,z) $ & $ \omega_{\rm tot} $ & $c_s^2 $ & Stability \\
		\hline
		$ P_{1\mp} $ & $ (\mp 0.945, \pm 2.56,0) $ & $ -1.63 $  & $ - 0.063 $ & Stable \\
		\hline
		$ P_{2\mp} $ & $ (\mp 1.08, \mp 1.53,0) $ & $ -0.57 $ & $  0.063 $ & Stable  \\
		\hline 
		$ Q_{\mp} $ & \makecell{ $\bigg(\mp \frac{\sqrt{-m^2+2 m+30}}{\sqrt{15}}, $ \\  $ \mp  \frac{30 \sqrt{-m^2+2 m+30}}{m^3-4 m^2-26 m+60} , $ \\ $ \pm \frac{\sqrt{15} \left(-2 m^2+5 m+28\right)}{\eta  (m-2)^2 \sqrt{-m^2+2 m+30}} \bigg)  $ \\ $ (\mp 1.43, \mp 3.61, \pm 9.31) $ }  & \makecell{$\frac{15 \left(-m^2+2 m+30\right)^2 \left(\frac{1}{15} \left(-m^2+2 m+30\right)-2\right)}{\left(m^3-4 m^2-26 m+60\right)^2}  $ \\ $\approx$ 0.33} &  \makecell{ $ m$ and $\eta $ \\ dependent.   \\ 0.49 } & Saddle \\
		\hline
		$ R_{\mp} $ & $ (\mp 1.41, \mp  2.74 ,0) $  & $ \approx 0 $  & $ 0.2 $  & Saddle \\
		\hline 
	\end{tabular}
	\caption{The critical points and their nature in presence of dust for $\delta = 20, $ for general $ m$ and $\eta$ (the top two rows), and for $ m= 1/2, \eta = 1$ (for the last two rows). }
	\label{tab:Model 01 matter critical points}
\end{table}

\subsubsection{Dynamics in presence of radiation}
In this subsection we will explore the interaction between $ k $-essence scalar field and radiation fluid. Using the same technique, as discussed above after Eq.~\eqref{F replacement}, we find the critical points $ P_{\mp} $ listed in Tab.~[\ref{tab:temp critical points model 01 radiation}]. We get the same $ \gamma -\delta $ relation shown in Fig.~[\ref{fig:gamma delta}] for these critical points. Hence choosing $ \delta = 20 $, we get three sets of critical points, which are tabulated in  
Tab.~[\ref{tab:Model 01 radiation critical points}] for any value of $ m $ and $\eta$. Out of these three sets of critical points, $ P_{1\mp} $ and $ P_{2\mp} $ are similar to the previous case. Only critical points $ Q_{\mp} $ are both $ m $ and $\eta$ dependent. At these points the grand EOS parameter depends on $ m $ and $\eta$. For $ m=-2/3 $ the points signify the matter phase. The sound speed limit puts further constraint on $\eta$ as $ \eta \le -0.19 $ or $ -0.03 < \eta < 1.7 $. Using all these mentioned constraints we have tabulated the critical points and their corresponding properties in Tab.~[\ref{tab:Model 01 radiation critical points}] for $ \delta = 20, m= -2/3$ and $\eta = 1$. 

\begin{figure}[t]
	\begin{minipage}[b]{0.5\linewidth}
		\centering
		\centering
		\includegraphics[scale=.5]{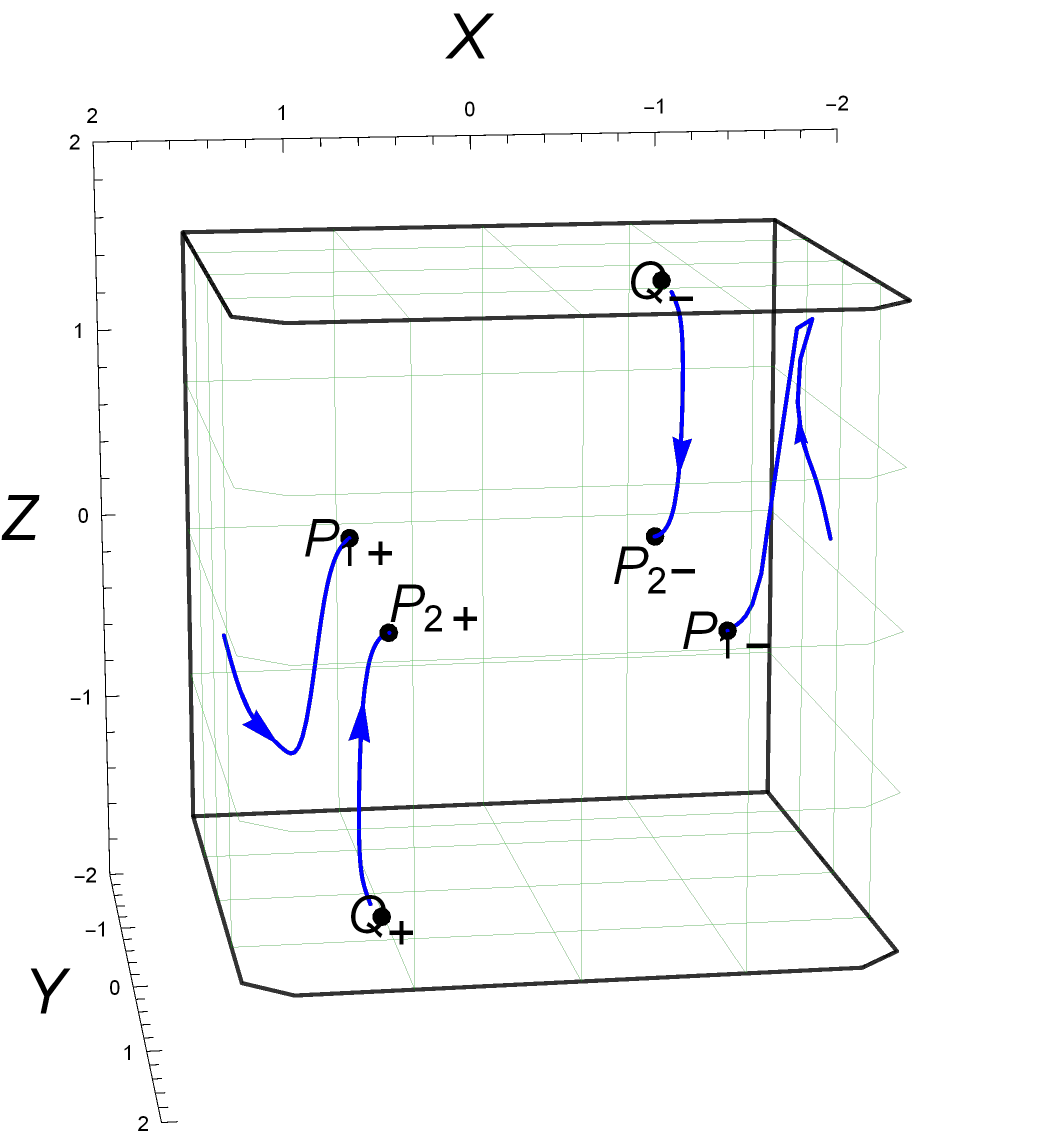}
		\subcaption{3D-Phase space for radiation dominated fluid with $\delta = 20, m=-2/3, \eta = 1$.}
		\label{fig:Model 01 radiation phase space}
	\end{minipage}
	\hspace{0.2cm}
	\begin{minipage}[b]{0.5\linewidth}
		\centering
		\includegraphics[scale=0.65]{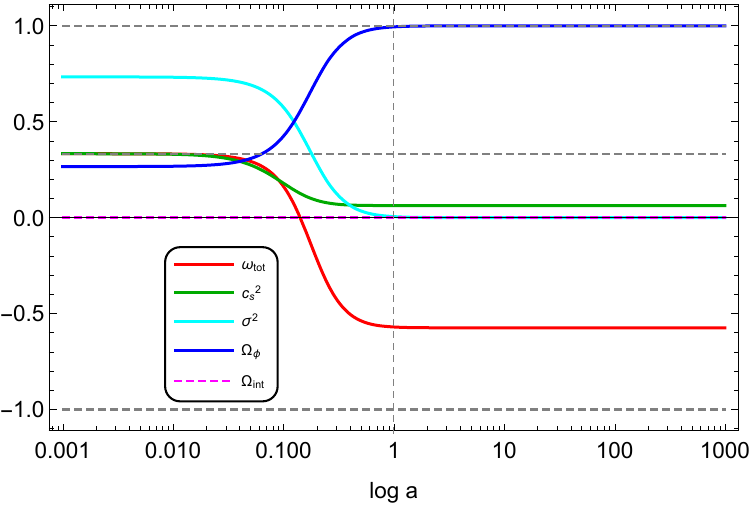}
		\subcaption{Evolution in the presence of radiation background fluid with $\delta = 20, m= -2/3$ and $ \eta = 1 $.}
		\label{fig:Model 01 radiation evolution}
	\end{minipage}
	\caption{3D Phase Space and Evolution plot in presence of radiation for Model I}
\end{figure}

From the phase space plot in Fig.~[\ref{fig:Model 01 radiation phase space}] (where again we have used the variables $X,Y,Z$ as defined in Eq.~(\ref{phasedef}))  we notice that both $P_{1\mp}$ and $P_{2\mp}$ are stable attractors. The trajectories nearby $ Q_{\mp} $ evolve towards the $ P_{2\mp} $, and makes the $ P_{2\mp} $ into a global attractor near which the universe shows accelerating expansion. This figure depicts that the universe may start at some point and gets attracted towards the matter phase $ Q_{\mp} $ and then eventually end up towards the accelerating phase. 

In Fig.~[\ref{fig:Model 01 radiation evolution}] the dynamics of the system becomes highly non-linear. We have plotted the regions where the grand EOS parameter of the system enters the radiation phase. During this phase, fluid density ($ \sigma^2 $) dominates and interaction energy density remains subdominant. As the $ k $-essence energy density evolves, the fluid density decreases, and the value of grand EOS parameters becomes zero. The change in phase is very steep, and at this phase, $ k $-essence energy density becomes greater than fluid density.  This signify that the field itself behaves like the dark-matter candidate. As $ k $-essence dominates in the late-time phase the system enters the accelerating phase. During the whole evolution period sound speed lies between $0$ to $1$.
\begin{table}[t]
	\tiny
	\begin{center}
		\centering
		\begin{tabular}{|c|c|c|c|}
			\hline
			\makecell{Critical \\ Points} & $ (x,y,z) $ & $ \omega_{\rm tot} $ & $ \gamma $\\
			\hline
			$ P_{\mp} $ & $ \left( \mp \frac{\sqrt{30 \gamma ^2 \delta +\gamma  (48-36 \delta )-64}}{3 \sqrt{\gamma  (3 \gamma -4) \delta }}, \mp \frac{3 \sqrt{\frac{3}{2}} \gamma  \sqrt{\delta } \sqrt{\gamma  (3 \gamma -4) \delta }}{2 \sqrt{15 \gamma ^2 \delta -18 \gamma  \delta +24 \gamma -32}},0 \right)  $ & $ \frac{1}{3}-\frac{\gamma ^2 \delta }{4} $& $  \dfrac{-6 \mp 2 \sqrt{12 \delta +9}}{3 \delta }$\\
			\hline
			$ P_{1\mp} $ & $ \left(\mp \frac{\sqrt{\frac{2 \sqrt{12 \delta +9}+\delta  \left(\sqrt{12 \delta +9}+7\right)+6}{\delta }}}{\sqrt{\frac{3 \sqrt{12 \delta +9}+\delta  \left(\sqrt{12 \delta +9}+9\right)+9}{\delta }}}, \pm \frac{\left(\sqrt{12 \delta +9}+3\right) \sqrt{\frac{3 \sqrt{12 \delta +9}+\delta  \left(\sqrt{12 \delta +9}+9\right)+9}{\delta }}}{\sqrt{3} \sqrt{\delta } \sqrt{\frac{2 \sqrt{12 \delta +9}+\delta  \left(\sqrt{12 \delta +9}+7\right)+6}{\delta }}},0 \right)  $ & $-\frac{2 \left(\sqrt{12 \delta +9}+3\right)}{3 \delta }-1 $ & $ -\frac{2 \left(\sqrt{12 \delta +9}+3\right)}{3 \delta } $ \\
			\hline 
			$ P_{2 \mp} $ & $ \left(\mp \frac{\sqrt{\frac{\delta  \left(7-\sqrt{12 \delta +9}\right)-2 \sqrt{12 \delta +9}+6}{\delta }}}{\sqrt{\frac{\delta  \left(9-\sqrt{12 \delta +9}\right)-3 \sqrt{12 \delta +9}+9}{\delta }}}, \mp \frac{\left(\sqrt{12 \delta +9}-3\right) \sqrt{\frac{\delta  \left(9-\sqrt{12 \delta +9}\right)-3 \sqrt{12 \delta +9}+9}{\delta }}}{\sqrt{3} \sqrt{\delta } \sqrt{\frac{\delta  \left(7-\sqrt{12 \delta +9}\right)-2 \sqrt{12 \delta +9}+6}{\delta }}},0 \right)  $ &$ \frac{-3 \delta +2 \sqrt{12 \delta +9}-6}{3 \delta } $ & $ \frac{2 \sqrt{12 \delta +9}-6}{3 \delta } $ \\
			\hline
			
		\end{tabular}
		\caption{Critical points, in presence of radiation, for general $ m $ and $ \eta $. }
		\label{tab:temp critical points model 01 radiation}
	\end{center}
\end{table}
\begin{table}[t]
	
	\centering
	\begin{tabular}{|c|c|c|c|c|}
		\hline
		\makecell{Critical \\ Points} & $ (x,y,z) $ & $ \omega_{\rm tot} $ & $c_s^2 $ & Stability \\
		\hline
		$ P_{1\mp} $ & $ (\mp 0.945, \pm 2.56,0) $ & $ -1.63 $  & $ - 0.063 $ & Stable \\
		\hline
		$ P_{2\mp} $ & $ (\mp 1.08, \mp 1.53,0) $ & $ -0.57 $ & $  0.063 $ & Stable  \\
		\hline 
		$ Q_{\mp} $ & \makecell{ $\left(\mp\frac{\sqrt{984 \eta +(3-18 \eta ) m^2+12 (2 \eta -1) m+172}}{4 \sqrt{30 \eta +15}}, \right.$\\  $ \mp  \frac{160 \sqrt{2 \eta +1} \sqrt{984 \eta +(3-18 \eta ) m^2+12 (2 \eta -1) m+172}}{(m-2) \left(-4 (246 \eta +43)+3 (6 \eta -1) m^2+(12-24 \eta ) m\right)}, $ \\ $\left. \mp \frac{16 \sqrt{\frac{5}{3}} \left(-3 m^2+6 m+80\right) \sqrt{984 \eta +(3-18 \eta ) m^2+12 (2 \eta -1) m+172}}{\sqrt{2 \eta +1} (m-2)^2 \left(-4 (246 \eta +43)+3 (6 \eta -1) m^2+(12-24 \eta ) m\right)}\right) $ \\, \\
			$ (\mp 1.26, \mp 3.08, \pm 3.71) $ }  & \makecell{$ m $ and $\eta$ \\ dependent \\ $\approx$ 0} &  \makecell{ $ m$ and  $\eta $ \\ dependent.   \\ 0.47 } & Saddle \\
		\hline
	\end{tabular}
	\caption{Model I critical points, in presence of radiation, for $\delta = 20, $ and general $ m$ and $\eta$, and stability for $ m= -2/3, \eta = 1$. }
	\label{tab:Model 01 radiation critical points}
\end{table}

\subsection{Study of Model II}

The chosen form of the interaction that corresponds to this model is $ f= \dfrac{F}{n \kappa^2 \phi^2} $, where $ F= X^2-X$ is the usual kinetic term of the $  k$-essence field. In this model the interaction term does not contain the energy density of the fluid $ \rho $; rather it only depends on one of the fluid parameters, the number density $ n $. Because of this the dimensionality of the phase space reduces from three to two. Therefore the interacting system can be studied using only two dynamical variables $ x $  and $ y $. The model parameters have been evaluated in Tab.~[\ref{table:M1}]. The constrain equation can be written as:
\begin{equation}\label{modelII constrained}
	\begin{split}
		\sigma^2	&  = 1 - y^2 x^2 \left[ (x^2 -1) \left(\dfrac{3}{4} - \dfrac{x}{\delta} \right) + \dfrac{1}{4} \right]\,. 
	\end{split}
\end{equation}
The equation of the state and generalized adiabatic sound speed of the system can be defined as:
\begin{equation}\label{}
	\omega_{\rm tot} = \omega \sigma^2 + y^2 F\left(1 + \dfrac{1}{\delta} x \right), \quad  c_{s}^2  = \dfrac{F_{,X} + x F_{,X}/\delta + F/(x \delta)}{(F_{,X} + 2 X F_{,XX}) + F_{,XX} x^3/\delta + 3 x F_{,X} /\delta} \,. 
\end{equation}
Using these quantities we study the dynamical development of the system.

\subsection{Dynamics of Model II}

In the present case the equation of state and the sound speed are not invariant under $ x \to -x  $ transformation. Sound speed is solely a function of dynamical variable $ x $. The critical points of the system can be found by solving the autonomous equations $ x' = 0, y' = 0$ in Eq.~\eqref{auto_x} and Eq.~\eqref{auto_y} for some values of $\delta$. The system turns out to be complicated and hence it is difficult to calculate analytically the critical points in terms of the model parameters. Hence we have adopted the numerical technique to find the critical points. We choose $\delta$ in such a way that at a late time it produces enough acceleration. Corresponding to the specific choice of $\delta$ we will find the dynamics of the system. We have found no critical points at infinity in the present case. 

\subsubsection{Dynamics in the presence of dust}

\begin{figure}[t]
	\begin{minipage}[b]{0.5\linewidth}
		\centering
		\includegraphics[scale=.4]{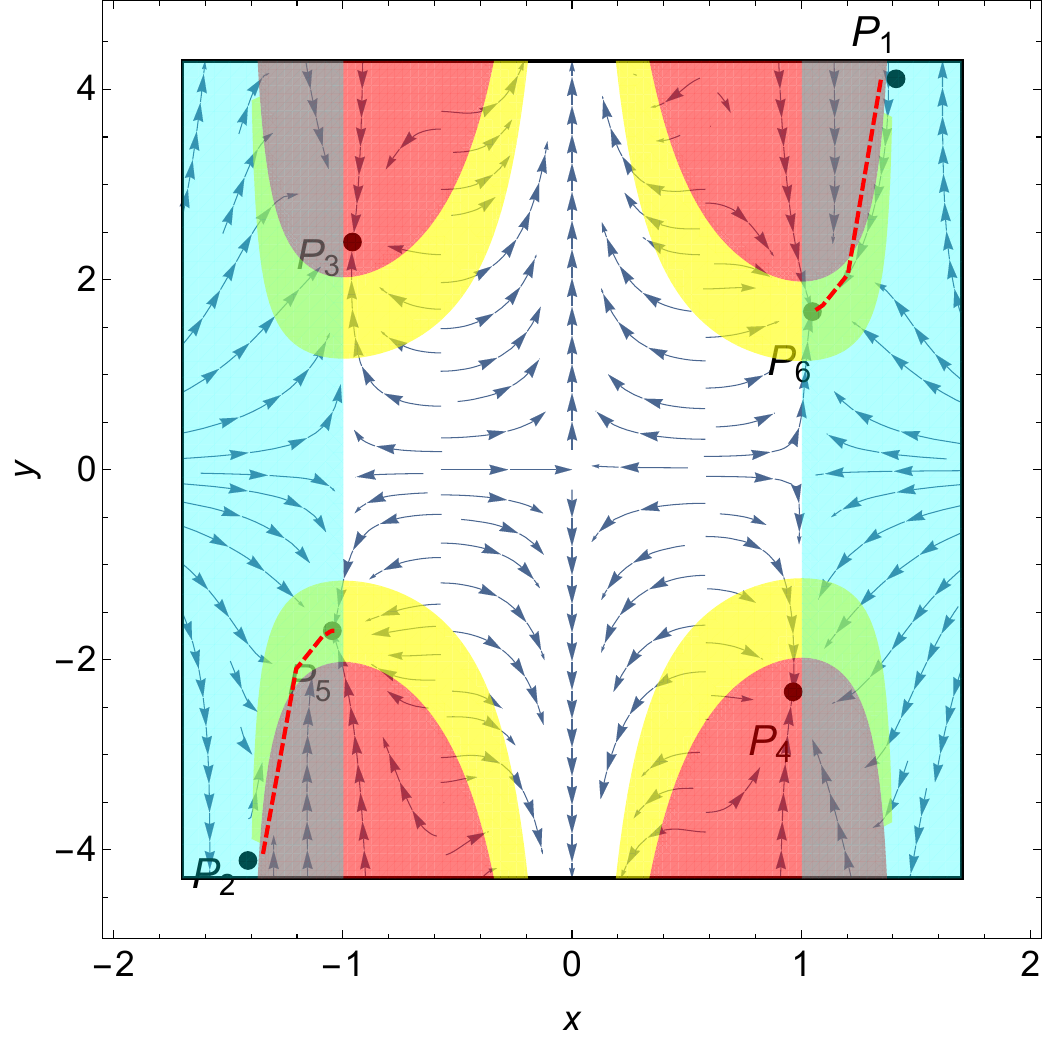}
		\subcaption{Model II Phase space in the presence of dust with $\delta=45.11$.  }
		\label{fig:Model 2 matter phase space}
	\end{minipage}
	\hspace{0.2cm}
	\begin{minipage}[b]{0.5\linewidth}
		\centering
		\includegraphics[scale=0.65]{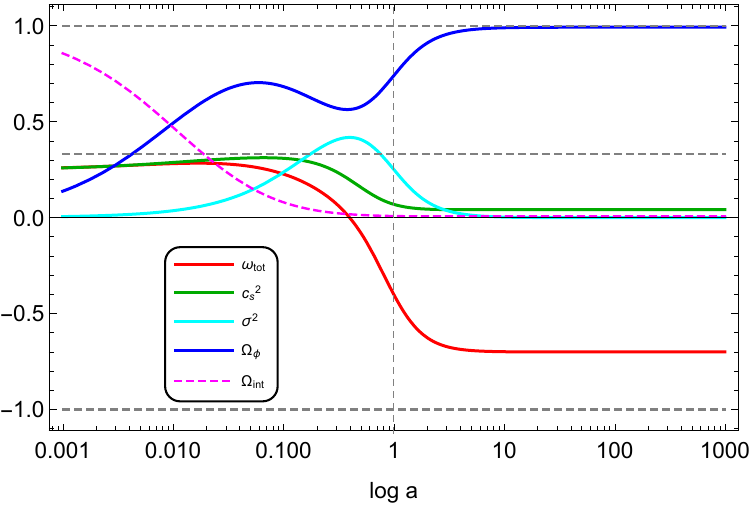}
		\subcaption{Evolution in the presence of dust with $\delta=45.11$.  }
		\label{fig:Model 02 matter evolution}
	\end{minipage}
	\caption{In phase space the blue region shows the sound speed limit $ 0<c_{s}^2<1 $, red region shows the phantom $ \omega_{\rm tot}< -1 $ and yellow region signifies $ - 1\le \omega_{\rm tot} \le -1/3 $.}
\end{figure}

We have found six critical points in the present case, when the fluid has $\omega=0$. The critical points are shown in Tab.~[\ref{tab:critical points model 02 delta45}]. Out of these six critical points, $ P_{1}, $ and $ P_{2} $ are saddle points and the corresponding grand EOS parameters near these points are approximately zero; hence these critical points signify the matter-dominated phase of the universe. At these phase points the sound speed is $\approx 0.2$. The other four critical points of this scenario are stable. To understand the nature of the critical points we have depicted the 2-D phase space, shown in Fig.~[\ref{fig:Model 2 matter phase space}]. The phase space is divided into red, green, and yellow regions, which signify the system's physical characteristics. The red region signifies the phantom phase, the yellow region corresponds to the accelerating phase and the blue region shows the the region where the sound speed limit is obeyed. The rest of the region shows a non-accelerating phase. The nearby trajectories of points $ P_{1} $ and $ P_{2} $ are repelled from these and attracted towards the points $ P_{6} $ and $ P_{5} $. A red phase space trajectory has been plotted by solving the autonomous system, which connects $ P_{1} $ to $ P_{6} $ and $ P_{2} $ to $ P_{5} $ by via heteroclinic orbits. The evolution of the trajectory from $ P_{1} $ and $ P_{2} $ signifies that the universe evolves from the matter dominated phase and enters into the accelerating phase. It turns out that $ P_{5} $ and $ P_{6} $ are globally attractors near which the sound speed remains small. The other fixed points $ P_{3} $ and $ P_{4} $ correspond to the phantom phase with a negative sound speed. The nearby vector points are attracted towards these points; hence the phantom phase of the system is stable. 

The dynamic evolution of the system is shown in Fig.~[\ref{fig:Model 02 matter evolution}]. It is seen that at an early epoch the grand EOS is $ \approx 0.25 $. At this phase both $ k $-essence energy density and fluid energy density $ \sigma^2 $ are subdominant while $\Omega_{\rm int}$ dominates. The early phase mimics a radiation dominated phase. As the universe evolves both $ k $-essence energy density and $\sigma^2$ increase and grand EOS decreases. When $\sigma^2$ increases the grand EOS becomes zero. At this point both interaction density and $\Omega_{\phi}$ are subdominant. It is seen that as $\Omega_{\phi}$ starts increasing the grand EOS decreases and saturates to $ \approx -0.7 $. At the late time both fluid density and interaction density become subdominant, and the sound speed of the universe becomes close to zero. In presence of dust, the coupling term explains the late-time scenario of the universe and it also gives rise to some interesting behavior in the early phase. 
\begin{table}[t]
	
	\centering
	\begin{tabular}{|c|c|c|c|c|}
		
		\hline
		\makecell{Critical \\ Points} & $ (x,y) $ & $ \omega_{\rm tot} $ & $ c_{s}^2 $ & Stability\\
		\hline
		$ P_{1} $ & $ (1.414,4.112) $ & $ -0.005$ & $ 0.198 $ & Saddle \\
		\hline
		$ P_{2} $ & $ (-1.414,-4.112)$ & $-0.097$ & $ 0.202$ & Saddle\\
		\hline 
		$ P_{3} $ & $ (-0.958,2.395)$  & $ -1.394$ & $ -0.043 $ & Stable\\
		\hline 
		$ P_{4} $ & $ (0.965, - 2.337) $ & $ -1.388 $ & $ -0.042 $ & Stable\\
		\hline
		$ P_{5} $ & $ (-1.046,-1.695) $ &$ -0.695 $ & $ 0.044 $ & Stable\\
		\hline
		$ P_{6} $ & $ (1.049,1.663) $ & $-0.70 $ & $ 0.041  $ & Stable \\
		\hline
		
	\end{tabular}
	\caption{The critical points for Model II, in the presence of dust, for $\delta = 45.11$. }
	\label{tab:critical points model 02 delta45}
\end{table}

\subsubsection{Dynamics in presence of radiation}

\begin{figure}[t]
	\begin{minipage}[b]{0.5\linewidth}
		\centering
		\includegraphics[scale=.4]{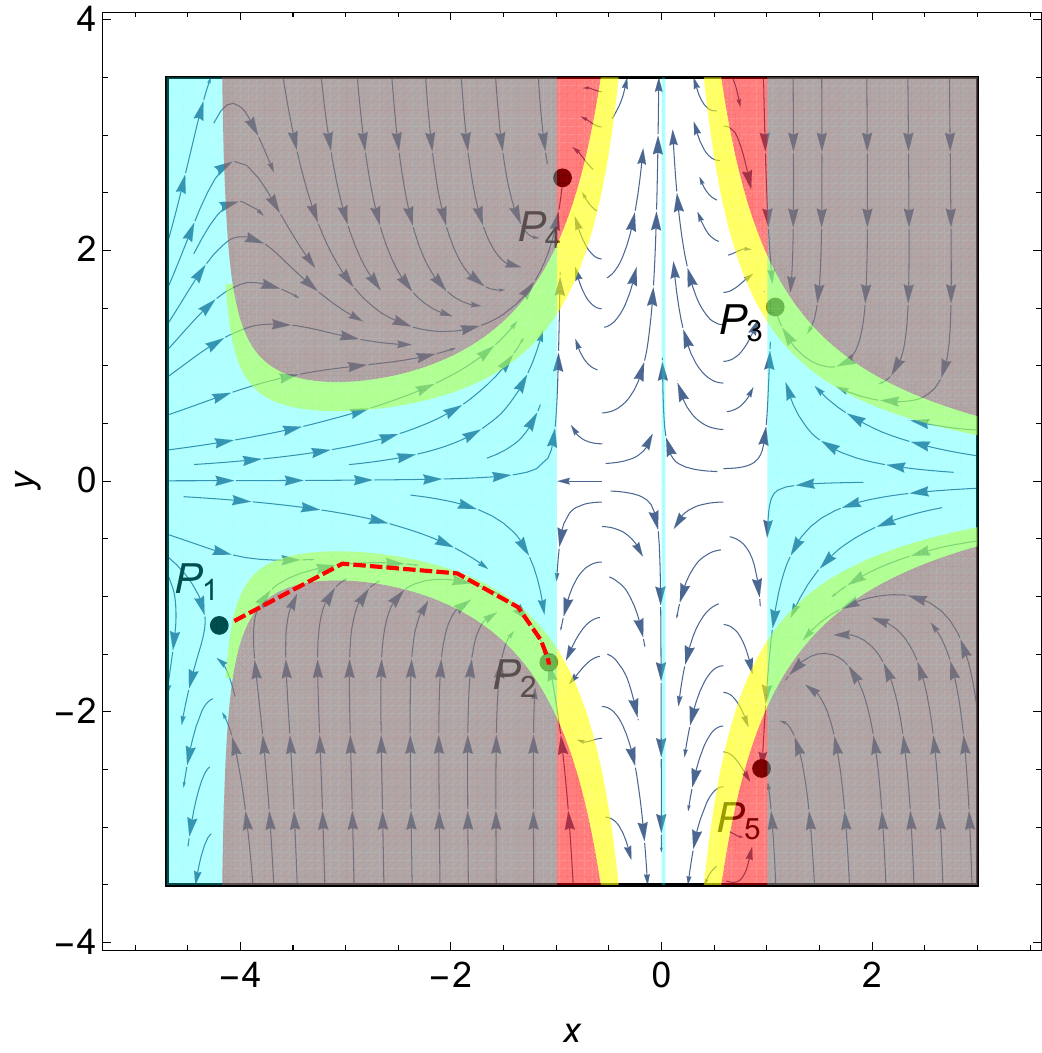}
		\subcaption{Model II Phase space dynamics in presence of radiation with $\delta = 20.63$.}
		\label{fig:Model 2 radiation phase space}
	\end{minipage}
	\hspace{0.2cm}
	\begin{minipage}[b]{0.5\linewidth}
		\centering
		\includegraphics[scale=0.65]{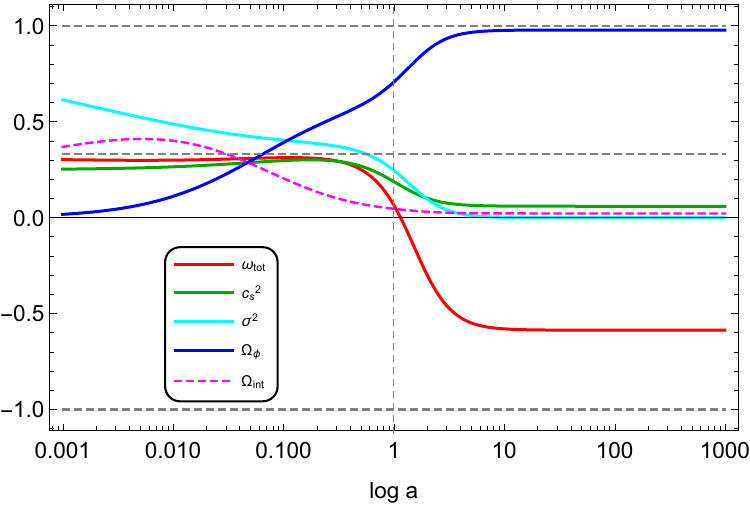}
		\subcaption{Evolution in the presence of radiation with $\delta = 20.63$.}
		\label{fig:Model 02 radiation evolution}
	\end{minipage}
	\caption{In phase space the blue region shows the sound speed limit $ 0<c_{s}^2<1 $, red region shows the phantom $ \omega_{\rm tot}< -1 $ and yellow region signifies $ - 1\le \omega_{\rm tot} \le -1/3 $.}
\end{figure}

In the present case we have chosen the model parameter $ \delta = 20.63 $. For this benchmark point we found a total of five fixed points out of which one of the fixed points $ P_{1} $ corresponds to radiation dominated phase. We have listed all the fixed points and their nature in Tab.~[\ref{tab:Model 2 radiation critical point delta 20}]. The 2-D phase space is displayed in Fig.~[\ref{fig:Model 2 radiation phase space}]. We see that $ P_{2} $ and $ P_{3} $ are fixed points near which we get accelerated expansion with positive sound speed and nearby trajectories are attracted towards them. The fixed point corresponding to the radiation phase is $ P_{1} $, some of the trajectories are attracted towards it and finally they get repelled. The evolution around this point is shown by a red trajectory which connects with $ P_{2} $ and completes a heteroclinic orbit. This shows that the radiation phase evolves into an accelerating phase. There are some nearby vector points of $ P_{1} $, which shows that they are repelled from this point and go towards $ P_{2} $ from the blue-white region, which may explain why the universe goes from radiation to matter phase and then finally enters into the accelerating phase. This point will become clear when we analyze the system dynamics. There exists no non-accelerating fixed point corresponding to $ P_{3} $, which signifies the non-symmetric nature of the autonomous system. Here $ P_{2} $ becomes the global attractor. Like the previous case, the phantom fixed points are stable with nearly zero sound speed. The dynamics as shown in Fig.~[\ref{fig:Model 02 radiation evolution}] has similar behavior compared to our previous case (dynamics in presence of dust) in late-time phase. However, in the early epoch, the fluid density  dominated with EOS $ 1/3 $. The interaction density increases for some time but then it decreases gradually.

In the present case the $ k $-essence density, $ \Omega_{\phi} $, is monotonically increasing and dominates at the late time. As the universe evolves from the early phase, the grand EOS parameter starts decreasing and crosses the $ 0 $ line, specifying a brief matter phase of the universe. This transition happens due to the dynamic characteristics of the $ k $-essence field. Hence $ k $- essence is solely responsible for the non-accelerated matter phase. During this transition the interaction density is non-zero but highly sub-dominated. Hence radiation background fluid coupled non-minimally with $ k $-essence can explain the early phase by the non-zero interaction between the two sectors, but at the late-time $ k $-essence explains the accelerating scenario the universe. 

\begin{table}[t]
	\centering
	\begin{tabular}{|c|c|c|c|c|}
		\hline
		\makecell{Critical \\ Points} & $ (x,y) $ & $ \omega_{\rm tot} $ & $ c_{s}^2 $ & Stability\\
		\hline
		$ P_{1} $ & $ (-4.20, -1.25) $ & $ 0.33	$ & $ 0.36$ & Saddle \\
		\hline
		$ P_{2} $ & $ (-1.07, -1.57)$ & $ -0.57 $ & $ 0.07 $ & Stable\\
		\hline 
		$ P_{3} $ & $ (1.08, 1.51)$  & $ -0.59$ & $0.06 $ & Stable\\
		\hline 
		$ P_{4} $ & $ (-0.94, 2.63) $ & $ -1.63 $ & $ -0.07 $ & Stable\\
		\hline
		$ P_{5} $ & $ (0.95, -2.49) $ &$-1.60 $ & $ -0.06$ & Stable\\
		\hline	
	\end{tabular}
	\caption{The critical points for Model II, in the presence of radiation, for $\delta = 20.63$. }
	\label{tab:Model 2 radiation critical point delta 20}
\end{table}
%

\section{Conclusion}
\label{Conclusion}

This paper explores the cosmological dynamics of a non-minimally
coupled $k$-essence scalar field and a relativistic fluid using a
variational approach.  The variational approach is a valuable
technique to handle the theories involving fields and fluids in a
covariant way. The main reason for choosing the variational approach
is that it can produce the dynamical equations of cosmological
development from a very simple premise. This approach can also explain
more radical type of interactions between a scalar field and matter
fluid. We have studied this field-fluid coupled system where the fluid
component is radiation or dust. The coupling term in the level of the
action has been chosen as $f(n,s,\phi,X)J^{\mu}\partial_{\mu} \phi$,
in which the interaction term $f$ depends on both field and fluid
parameters. We have explicitly coupled the space-time derivative of
the field $\phi$ with the fluid 4-velocity $ U^{\mu} $ which gives
rise to a highly non-linear dynamical evolution. The non-linear
effects are amplified in this model because for non-zero interaction
energy density, any change in the field value immediately affects the
fluid sector via the derivative coupling term. If the field energy
density is rapidly changing in some phase then the fluid properties
will also rapidly change in that phase.  On the other hand if the
field value does not change with time this kind of interaction
vanishes.  Due to the presence of the specific derivative dependent
interaction term several thermodynamic quantities such as system's
temperature $ T_{\rm tot} $ and chemical-free energy $ F_{\rm tot} $
have been generalized. We have established that the total energy
momentum of the field-fluid coupled system is conserved. The other
matter quantities like the equation of state (EOS) and sound speed ($
c_{s}^2 $) have been generalized properly.

{The existing cosmological observational results are generally
compatible with the $\Lambda$-CDM model, however, there have been
disagreements on the value of the Hubble constant $H_0$ in the high
and low-redshift regions. In this regard, the results of several dark energy theories closely match
with the observational data sets and this matching turns out to be better than the matching of observational data with $\Lambda$-CDM model
results. The resolution of the tension related to the mismatch of the
values of $H_0$ in the high and low-redshift measurements, however,
remains a difficult problem. Interacting dark sector is one of the
various possibilities which can address this issue. Previous authors
have tried several phenomenological approaches to the interacting dark
energy scenario in which the coupling in the dark sector is
introduced, by hand, at the level of the continuity equations (for energy densities). As
these interactions are not introduced at the level of the action these
theories may give rise to instabilities at the perturbation level, as
discussed in Ref.~\cite{Tamanini:2015iia}.  In the present work, we construct
a coupled CDM scenario where the coupling is introduced at the Lagrangian level and consequently all the
physical parameters can be consistently defined in the background and perturbation
levels.  From recent data analysis based on Planck result and Hubble
Space Telescope result, some theorists do think that a minimally
coupled dark energy and CDM may not effectively reduce the $H_0$
tension. A recent study on interacting dark matter with dark energy
showed that non-minimal coupling in the dark sector can significantly
reduce the $H_0$ tension \cite{DiValentino:2019ffd}. Following the
success of interacting dark sector, we have tried to build a 
generalized model of derivative-dependent coupling of the field and fluid sectors
which relies on robust mathematical foundations. The physical
significance of the presented models can be testified in the future
when our predictions are matched with observational results.

}

We have constructed a dynamical system to harness the dynamics of the coupled field-fluid system. The primary variables have been selected as $ x $, $ y $, $\sigma$ and $ z $. Due to the presence of a constraint the dynamical system has three dimensional phase space. To compactly write the dynamical equations we have defined some other parameters. The other parameters are expressed in terms of the variables $ x $, $ y $, $\sigma$ and $ z $. Hence in general we have constructed the 3-D autonomous system which encodes all the information regarding the system. To explore the dynamics we have constructed two models, Model I and Model II using the  inverse square potential $V(\phi) \propto 1/\phi^2$. In all the cases the the $k-$essence kinetic term $F(X)=X^2-X$. To work out the dynamics we have used the simplest models of interaction although these simplest models do contain multiple constant parameters. There can be other models of interaction which may contain more constant parameters. 

In Model I we studied the system in presence of dust and radiation separately. It  was seen that the critical points had various dependencies on the model parameters $\delta$, $m$ and $\eta$. Some of the critical points are only dependent on $\delta$. Those critical points were found out in a heuristic manner. Our method  showed that some critical points lie in the phantom domain and others lie in the non-phantom region. We used the grand EOS and the sound speed limit to select other the constant model parameters. In presence of dust we have found the critical points that correspond to different phases of the universe. These critical points specify various phases of the universe, from radiation domination to accelerating and phantom phases. Due to the presence of the derivative dependent interaction term, the interaction energy density and scalar field density plays important role in the early phase of the universe, whereas scalar field density dominates over other components in the late phase. In presence of radiation, due to the interaction, the early phase becomes highly non-linear, and hence most  physical quantities increase abruptly. As the system evolves and enters the radiation phase the fluid density dominates. The transition from the radiation to matter phase is abrupt and the system slowly enters into the late-time cosmic acceleration phase.

In  Model II the system reduces from a 3-D dynamical system into a 2-D system governed by $ x $ and $ y $ only. When the fluid present corresponds to dust we have selected the model parameter $\delta = 45.11$, for which we found a set of critical points which can describe the evolution of the universe from the matter dominated phase to the phantom phase of the universe. We found a heteroclinic orbit in phase space which shows how the matter dominated phase evolves and is finally attracted towards a stable accelerating point. The phantom points are also stable, but we have not found any critical points corresponding to the early phase which connects to them. Hence these phantom fixed points become isolated. In the early phase the interaction density dominates and the system is in a radiation dominated phase. While at the matter phase, fluid density becomes dominant, and the field density dominates the late-time phase. 

In presence of radiation the phase space has five critical points, out of which only one critical point shows the radiation epoch. This point evolves towards the accelerating point and makes a  heteroclinic orbit. Other critical points become isolated stable points. We see from the evolution plot that the fluid density becomes dominant at an early epoch which signifies the radiation phase, and at some point, interaction density also increases and then decreases. So in the presence of radiation the behavior of the system becomes a little interesting in the early epoch but then the system enters an accelerating phase after a brief matter epoch. Hence from both models it becomes vivid that the late-time dynamics are solely played by the $ k $-essence scalar field, but as we move towards the early phases, the non-linear effects of interaction play a more significant role. We see that the field-fluid model collectively explains different epochs of the universe, from radiation to accelerating and phantom epochs.

{ In the previous study with algebraic coupling in the dark sector\cite{Chatterjee:2021ijw} the authors showed that one can attain an accelerating phase of the universe in the late phase of cosmic evolution. In the present work, we have worked with the derivative coupling in the dark sector and obtained results which can be used for the late cosmic acceleration phase. In such a circumstance a comparison between the previous work\cite{Chatterjee:2021ijw} and the present work can be useful. First of all, we specify that both the works have somethings in common, namely the form of $F(X)$ and the form of the potential function $V(\phi)$ remains the same. There is another similarity between these two works, both aim to produce a stable, accelerating expansion phase in the future. We will show that the last point is obtained in two different ways in these two attempts. Except for these similarities, there are important differences at the level of the models. The forms of the functions, $f$, which accompany the couplings are functionally different in the two works and the parameters in $V(\phi)$ are also chosen differently so that the different models yield proper cosmic evolution. The number and nature of the various critical points of the systems are affected by the nature of the different types of couplings used in the two works. The qualitative difference between the nature of the critical points primarily distinguishes the present paper from the previous one. The previous study focused mainly on the dark sector whereas the present work includes non-minimal coupling of the dark energy sector with radiation. One must note that in the previous work, the late phase of the universe permits the non-minimal coupling in the dark sector whereas in the present case as the scalar field value saturates near the stable critical point, specifying the late phase of the universe, the non-minimal coupling ceases to play any significant role. In the final phase of the universe, dark energy and dark matter are two separate entities in the present models as the interaction is directly proportional to the rate of change of the scalar field. This is another important difference between these two attempts related to different couplings in the dark sector. There are many other differences between the two models based on two different couplings, some of these points are related to formal theoretical issues while others are related to the nature of cosmological dynamics. We end this discussion with one prominent difference in the level of cosmological dynamics which can have observational effects. In Model II of the previous work, the evolution of the effective (or total) EOS and $c_s^2$ show jumps before the final stable phase is obtained. These jumps specify a phase transition period. In the present work, the same parameters smoothly enter the late phase and no jumps or kinks are noticed. This fact may have interesting cosmological effects.}

\paragraph{Acknowledgement}
Authors are thankful to the referee for the valuable comments. A.C would like to thank Indian Institute of Technology, Kanpur for supporting this work by means of Institute Post-Doctoral Fellowship \textbf{(Ref.No.DF/PDF197/2020-IITK/970)}.

\appendix
\renewcommand{\theequation}{A.\arabic{equation}}
\setcounter{equation}{0}
\section{Conservation Equation of Fluid plus Interacting}\label{proof01}
Using the forms of $h_{\mu\nu}$ and $\ti{T}^{\lambda \nu}$ defined in section \ref{Conservation sectors} we can write
\begin{equation}\label{}
	\begin{split}
		h_{\m}\nb_{\lambda}\ti{T}^{\lambda \nu} &= h_{\m}\nb_{\lambda} \left[ (\tilde{\rho} + \tilde{p})U^\lambda U^\nu + \tilde{p}g^{\la}\right]\\
		& = h^{\lambda}_{\mu}\nb_{\lambda}\ti{p} + g_{\m}\nb_{\lambda}\left[ (\tilde{\rho} + \tilde{p})U^\lambda U^\nu\right] + U_{\mu}U_{\nu}\nb_{\lambda}\left[ (\tilde{\rho} + \tilde{p})U^\lambda U^\nu\right]  \\
		& = h^{\lambda}_{\mu}\nb_{\lambda}\ti{p} + \left[\nb_{\lambda}(\tilde{\rho} + \tilde{p}) U^{\lambda}U_{\mu} + (\tilde{\rho} + \tilde{p})(U_{\mu}(\nb_{\lambda}U^{\lambda})+ g_{\m}U^{\lambda}\nb_{\lambda}U^{\nu})\right] +\\ 
		& \left[ -\nb_{\lambda}(\tilde{\rho} + \tilde{p})U^{\lambda}U_{\mu} +(\tilde{\rho} + \tilde{p})(-U_{\mu}\nb_{\lambda}U^{\lambda} ) \right]\,.
	\end{split}
\end{equation}
Here $ \ti{\mu} = \mu = \p \rho / \p n  $ and $ \ti{p} = n\, \p \rho / \p n - \rho - n^2f_{,n} U^\alpha \p_{\alpha}\phi $ . Assuming $\gamma = n^2f_{,n} U^\alpha \p_{\alpha}\phi$ and using the fact that $\tilde{\rho}=\rho$  we have $\ti{\rho} + \ti{p} = n \mu - \gamma $. Using these facts we can now proceed as: 
\begin{equation}\label{stress tensor 02}
	\begin{split}
		h_{\m}\nb_{\lambda}\ti{T}^{\lambda \nu} &= h^{\lambda}_{\mu}\nb_{\lambda}\ti{p} + \left[\ (\tilde{\rho} + \tilde{p})(g_{\m}U^{\lambda}\nb_{\lambda}U^{\nu})\right] \\
		& = h^{\lambda}_{\mu}\nb_{\lambda}\ti{p} + (n\ti{\mu}U^{\lambda}\nb_{\lambda}U_{\mu}) - n^2f_{,n} U^\alpha \p_{\alpha}\phi (U^\lambda \nb_{\lambda}U_\mu)\\
		%
		& = h^{\lambda}_{\mu}\nb_{\lambda}\ti{p} +n\left[ 2U^{\lambda}\nb_{\big[\lambda}(\ti{\mu}U_{\mu\big]})- \nb_{\mu}\ti{\mu} - U^{\lambda}U_{\mu}\nb_{\lambda}(\ti{\mu})\right]\\
		& - \gamma (U^\lambda \nb_{\lambda}U_\mu)  \\
		%
		& = n\bigg[ 2U^{\lambda}\nb_{\big[\lambda}(\ti{\mu}U_{\mu\big]})-(\nb_{\mu}\dfrac{\p \ti{\rho}}{\p n}) - \dfrac{1}{n}h^{\lambda}_{\mu}\nb_{\lambda} \ti{\rho} + \dfrac{1}{n}(\delta^{\lambda}_{\mu}+U^{\lambda}U_{\mu}) \nb_{\lambda}\left( n\dfrac{\p \ti{\rho}}{\p n}\right) - \\
		& U^{\lambda}U_{\mu}\nb_{\lambda}\dfrac{\p \ti{\rho}}{\p n}                 \bigg]  - \gamma (U^\lambda \nb_{\lambda}U_\mu) - h^\lambda _\mu \nb_\lambda \gamma \\
		&= n\bigg[ 2U^{\lambda}\nb_{\big[\lambda}(\ti{\mu}U_{\mu\big]})- \dfrac{1}{n}\nb_{\mu}\ti{\rho} - \dfrac{1}{n}U^{\lambda}U_{\mu}\left(\dfrac{\p\ti{\rho}}{\p n}\nb_{\lambda}n +\dfrac{\p\ti{\rho}}{\p s}\nb_{\lambda}s \right)   + \dfrac{1}{n}\bigg(\dfrac{\p \ti{\rho}}{\p n} \nb_{\mu}n +\\ &  U^{\lambda}U_{\mu} (\nb_{\lambda}n) \dfrac{\p \ti{\rho}}{\p n}  \bigg)               \bigg] - \gamma (U^\lambda \nb_{\lambda}U_\mu) - h^\lambda _\mu \nb_\lambda \gamma\,.
\nonumber                
\end{split}
\end{equation}
As $s$ is a constant, the above equation can be written as
\begin{eqnarray}
h_{\m}\nb_{\lambda}\ti{T}^{\lambda \nu} = 2n\,U^{\lambda}\nb_{\big[\lambda}(\ti{\mu}U_{\mu\big]}) - \gamma (U^\lambda \nb_{\lambda}U_\mu) - h^\lambda _\mu \nb_\lambda \gamma\,.
\end{eqnarray}
Now let us first evaluate $ 2 n U^{\lambda}\nb_{\big[\lambda}(\ti{\mu}U_{\mu\big]} $ by  using Eq.~\eqref{jeqn}. We get  
\begin{equation}\label{this one}
	\ti{\mu}U_{\mu} = -\left( \varphi_{,\mu} +s \theta_{,\mu} +\beta_{A} \alpha_{,\mu}^{A}\right) -\left[ \left(- n\dfrac{\p f}{\p n} \right) U^{\nu}U_{\mu} + f \delta _{\mu}^{\nu} \right] \p_{\nu}\phi\,.
\end{equation}
Computing $ 2 n U^{\lambda}\nb_{\big[\lambda}(\ti{\mu}U_{\mu\big]} $ with the first half of above Eq.~\eqref{this one} we have:
\begin{equation}\label{}
	\begin{split}
		U^{\lambda}\nb_{\big[\lambda}(\ti{\mu}U_{\mu\big]})& =	U^{\lambda}\nb_{\big[\lambda}\left[ \nb_{\mu\big]}\varphi + s\,\nb_{\mu\big]} \theta + 	\beta_{A}\nb_{\mu\big]} \alpha^{A}			\right]  \\
		& = U^{\lambda}\left[\nb_{\big[\lambda} \nb_{\mu\big]}\varphi + s\, \nb_{\big[\lambda}\nb_{\mu\big]}\theta + \nb_{\big[\lambda} \beta_{A} \nb_{\mu\big]} \alpha^{A}\right] \\
		& = U^{\lambda}\left[\nb_{\lambda}\beta_{A}\nb_{\mu}\alpha^{A} - \nb_{\mu}\beta_{A}\nb_{\lambda}\alpha^{A} \right] \\
		& =0\,.
	\end{split}
\end{equation}
Since, $ \nb_{\big[\lambda}\nb_{\mu\big]} =\dfrac{1}{2}\left(  \nb_{\lambda} \nb_{\mu} - \nb_{\mu} \nb_{\lambda} \right)$ the covariant derivatives commute for torsion-less space times when acted on scalars such as $\varphi$ or $\theta$. Now the second part of Eq.~\eqref{this one} yields: 
\begin{equation}\label{}
	\begin{split}
		2n U^{\lambda} \nb_{\big[\lambda} \left( nf_{,n} U^\nu U_{\mu\big]} \p_\nu \phi -f \p_{\mu\big]} \phi \right) 
		=2 \times \dfrac{1}{2}nU^\lambda \big[ \nb_{\lambda} \left( nf_{,n} U^\nu (\p_\nu \phi) U_\mu - f \p_\mu \phi\right)  \\ -  \nb_\mu \left( nf_{,n} U^\nu (\p_\nu \phi) U_\lambda - f \p_\lambda \phi \right)   \big] \\
		= nU^\lambda \big[ \nb_\lambda  (nf_{,n} U^\nu \p_\nu \phi)U_\mu + (\nb_\lambda U_\mu )nf_{,n} U^\nu \p_\nu \phi - (\nb_\lambda f) \p_\mu \phi - f \nb_\lambda(\nb_\mu \phi) \\
		-\nb_\mu  (nf_{,n} U^\nu \p_\nu \phi)U_\lambda - (\nb_\mu U_\lambda )nf_n U^\nu \p_\nu \phi + (\nb_\mu f) \p_\lambda \phi + f \nb_\mu(\nb_\lambda \phi)
		\big]\\
		= 
		\big[\nb_\lambda  (nf_{,n} U^\nu \p_\nu \phi)n U^\lambda U_\mu + n U^\lambda (\nb_\lambda U_\mu )nf_{,n} U^\nu \p_\nu \phi - \\ \left\lbrace f_{,n} \nb_\lambda n   + f_{,X} \nb_{\lambda} X \right\rbrace  nU^\lambda \p_\mu \phi
		+n \nb_\mu  (nf_{,n} U^\nu \p_\nu \phi) + \left\lbrace f_{,n} \nb_\mu n + f_{,X} \nb_\mu X  \right\rbrace  nU^\lambda \p_\lambda \phi
		\big]\,.	
	\end{split}
\end{equation}
Inserting the above expression in Eq.~\eqref{stress tensor 02} we get:  
\begin{equation}
	\begin{split}
		h_{\m}\nb_{\lambda}\ti{T}^{\lambda \nu}	 = 
		\big[\nb_\lambda  (nf_{,n} U^\nu \p_\nu \phi)n U^\lambda U_\mu + n U^\lambda (\nb_\lambda U_\mu )nf_{,n} U^\nu \p_\nu \phi - \\ \left\lbrace f_{,n} \nb_\lambda n   + f_{,X} \nb_{\lambda} X \right\rbrace  nU^\lambda \p_\mu \phi
		+n \nb_\mu  (nf_{,n} U^\nu \p_\nu \phi) + \left\lbrace f_{,n} \nb_\mu n + f_{,X} \nb_\mu X  \right\rbrace  nU^\lambda \p_\lambda \phi
		\big] \\
		- n^2f_{,n} U^\alpha \p_{\alpha}\phi (U^\lambda \nb_{\lambda}U_\mu) - h^\lambda _\mu \nb_\lambda \left(n^2f_{,n} U^\alpha \p_{\alpha}\phi \right)\,.
        \end{split}
\nonumber        
\end{equation}
One can plug this result back in Eq.~(\ref{appeqn}) to verify the energy-momentum conservation condition. 

%


\end{document}